\documentclass{article}

\PassOptionsToPackage{table}{xcolor}

\usepackage[main,final]{neurips_2026}

\usepackage[utf8]{inputenc}
\usepackage[T1]{fontenc}
\usepackage{times}
\usepackage{inconsolata}                 
\usepackage{microtype}

\usepackage{amsmath}
\usepackage{amssymb}
\usepackage{mathtools}
\usepackage{amsthm}
\usepackage{nicefrac}
\usepackage{latexsym}

\usepackage{xcolor}                      
\usepackage{graphicx}
\usepackage{float}
\usepackage{wrapfig}
\usepackage{caption}
\usepackage{subcaption}
\usepackage{booktabs}
\usepackage{multirow}
\usepackage{siunitx}

\usepackage[most]{tcolorbox}
\usepackage{listings}
\usepackage{enumitem}
\usepackage{xspace}
\usepackage{url}
\usepackage[textsize=tiny]{todonotes}

\usepackage{hyperref}
\hypersetup{hidelinks}                   
\usepackage{algorithm}                   
\usepackage{algpseudocode}
\usepackage[capitalize,noabbrev]{cleveref}  

\setcitestyle{numbers,square,comma}

\theoremstyle{plain}

\theoremstyle{definition}

\theoremstyle{remark}

\newcommand{\name}{\textsc{CTM-AI}}

\newcommand{\xhdr}[1]{{\noindent\bfseries #1}.}
\newcommand{\valerr}[2]{\ensuremath{#1}{$_{\pm #2}$}}
\newcommand{\ctmrow}{\hspace{0.8em}\ +~\name}

\definecolor{rowgray}{gray}{0.94}        
\definecolor{ctmrowcolor}{gray}{0.92}
\definecolor{bestgray}{gray}{0.78}       

\title{\name: Evolving Multi-Agent Intelligence\\Inspired by a Model of Consciousness}

\author{%
  Haofei Yu$^{1}$\thanks{Equal contribution.}\quad
  Yining Zhao$^{1}$\footnotemark[1]\quad
  Lenore Blum$^{2}$\quad
  Manuel Blum$^{2}$\quad
  Paul Pu Liang$^{3}$\\[2pt]
  $^{1}$UIUC \quad $^{2}$CMU \quad $^{3}$MIT\\
  {\small\texttt{haofeiy2@illinois.edu}}%
}

\begin{document}

\maketitle

\vspace{-2mm}
\begin{abstract}
Despite remarkable advances, today's AI systems remain narrow in scope, falling short of the adaptive and continual intelligence that characterizes human capabilities. This gap has fueled longstanding debates about whether AI might one day achieve human-like generality or even consciousness. Flipping the debate, we present a concrete blueprint for a multi-agent AI system inspired by a formal machine model of consciousness called the Conscious Turing Machine (CTM).
Together, \name\ contains a large collection of powerful processors ranging from specialized experts (e.g., language models, multimodal models, tools, and APIs) to unspecialized general-purpose learners poised to develop their own expertise.
Crucially, for whatever problem must be dealt with, information from many processors is competitively selected, integrated, and exchanged appropriately to solve the task. Feedback mechanisms in \name\ enable the system to iteratively evolve its internal connections and communication. \name\ achieves state-of-the-art accuracy on MUStARD (72.28) and UR-FUNNY (72.13), outperforming multimodal and multi-agent frameworks. On tool-using and agentic tasks, \name\ achieves 10+ points of improvement on StableToolBench and WebArena-Lite. Overall, \name\ offers a principled and testable multi-agent AI system inspired by a model of consciousness. The code is available at \url{https://github.com/consciousness-lab/ctm-ai}.

\end{abstract}

\section{Introduction}


\begin{wrapfigure}{r}{0.45\linewidth}
    \centering
    \vspace{-12mm}
    \includegraphics[width=\linewidth]{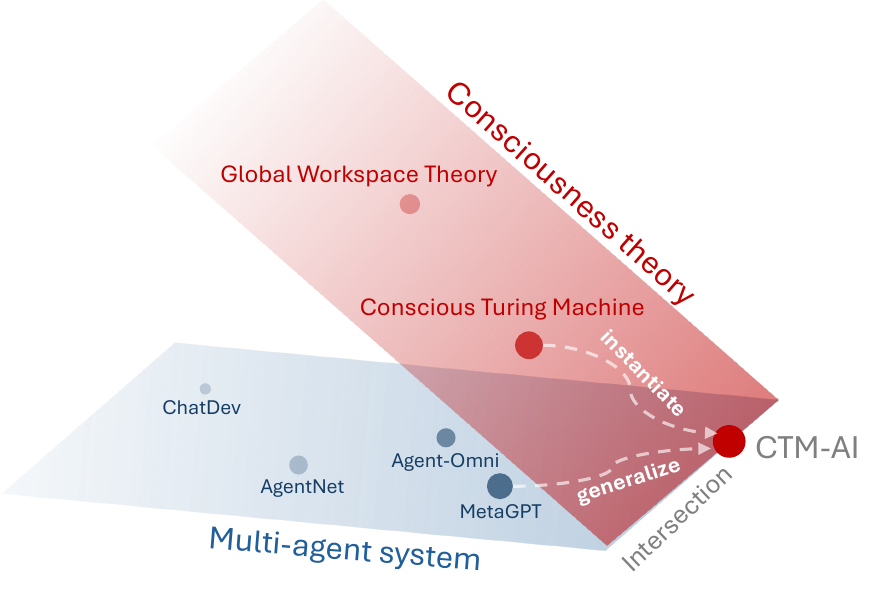}
    \caption{\small \textbf{Positioning of \name.} Existing research falls into two domains: models of consciousness (red) offer theoretical foundations but lack practical implementations, while multi-agent frameworks (blue) offer practical implementations but lack principled foundations. \name\ bridges this gap by \emph{instantiating} the Conscious Turing Machine as a practical AI system.}
    \label{fig:firstpage}
    \vspace{-7mm}
\end{wrapfigure}

In recent years, progress toward AI models capable of human-like intelligence has inspired debates regarding whether today's AI and its future counterparts can one day display human-like levels of consciousness. Flipping the debate, we present a concrete blueprint for a multi-agent AI system based on a formal machine model of consciousness, the Conscious Turing Machine (CTM)~\citep{blum2022theory,blum2021theory}. The CTM is a simple and formal model of consciousness inspired by Alan Turing’s model of computation~\citep{turing1936computable} and Bernard Baars’ theater model of consciousness~\citep{baars1993cognitive}. Critically different from other cognitive architectures and multi-agent LLM workflows, the CTM has no central executive – no conductor, no stage director~\citep{blum2023agi}. Instead, the CTM employs a global workspace and distributed competition to integrate the power of an enormous collection of parallel independent cognitive, sensory, motor, and extended long-term memory processors. When a problem needs to be solved, it becomes globally broadcast to all processors, eliciting help from those who might have the expertise, interest, and resources to tackle the problem, even though their talents and abilities might be unknown to a central executive.

While the CTM provides a theoretical foundation that is fully decentralized, modality agnostic, and architecturally modular, key mechanisms, including how processors broadcast information, form links, and combine expertise, and learn from feedback, are left abstract with no empirical validation. This raises a natural question: \emph{Can the theoretical architecture of the CTM be instantiated into a practically working multi-agent AI system?} To answer this question, our work bridges the gap between theory and practice by instantiating the formal CTM model as a multi-agent framework, \name, operationalized with modern foundation models. Specifically, \name\ features: (1) multiple specialized, agentic processors operating in parallel, (2) a limited-capacity short-term memory workspace that enforces selective attention via up-tree competition, (3) down-tree global broadcasting that distributes information from the workspace to all processors, and (4) iterative evolution through dynamic communication linking between processors, enabling integration of distributed knowledge into higher-order representations.

\xhdr{Key contributions} The significance of \name\ is threefold: (1) it serves as the first practical instantiation of the CTM, translating a theoretical model of consciousness into an executable multi-agent AI system; (2) it naturally yields a highly modular and decentralized architecture where processors can be flexibly added or removed, free from the rigid orchestrators of current agentic workflows; and (3) it enables iterative multi-agent evolution through decentralized collaboration and adaptive link formation, expanding reasoning and agentic capabilities beyond static architectures.

\xhdr{Main results} To evaluate \name's ability to coordinate multiple processors, modalities, and tasks, we present quantitative results across multimodal perception, tool use, and agentic environments. These benchmarks require systems to utilize external APIs, integrate and reason over multimodal data, and solve complex, multi-step problems. \name\ achieves state-of-the-art accuracy on MUStARD (72.28\%) and UR-FUNNY (72.13\%), outperforming unified multimodal baselines and multi-agent frameworks like MoA and MetaGPT. Furthermore, \name\ generalizes to tool-use and agentic tasks, yielding an absolute improvement of >10\% in pass rate on both StableToolBench and WebArena-Lite. Finally, our analysis demonstrates that: (1) \name\ can iteratively self-improve by adapting its processor links based on task complexity; (2) it integrates seamlessly with existing reasoning paradigms; and (3) its core dynamics are robust to hyperparameters and not over-engineered.

\section{Related Works}

\xhdr{Models of consciousness}
Computational models of consciousness seek to formalize how the human brain selects, integrates, and distributes information~\citep{butlin2023consciousness}. Alongside Global Workspace Theory (GWT)~\citep{baars1993cognitive}, several models have been developed with distinct emphases, including Integrated Information Theory~\citep{tononi2004information}, Higher-Order Theories~\citep{rosenthal2005consciousness}, and cognitive architectures such as ACT-R~\citep{anderson1997act} and SOAR~\citep{laird1987soar}. Within the GWT lineage, LIDA~\citep{franklin2013lida} implements GWT's cognitive cycle symbolically, the Global Neuronal Workspace~\citep{mashour2020conscious} formalizes it at the neural level, and the Global Latent Workspace~\citep{vanrullen2021deep} proposes a deep-learning roadmap for GWT-style integration. The Conscious Turing Machine (CTM)~\citep{blum2022theory,blum2021theory} formalizes GWT in the framework of Turing computation, defining mechanisms including up-tree competition for workspace access and down-tree broadcast~\citep{blum2023agi}, but has remained purely theoretical. \name\ provides a finite-scale operationalization of the CTM's competition, broadcast, and link-formation mechanisms using modern AI technologies.

\xhdr{Multi-agent frameworks}
Most recent multi-agent systems~\citep{kim2025towards,schmidgall2025agent} share two structural properties: a central executive that orchestrates information flow (e.g., the manager in MetaGPT), and task-specific workflows where each agent is bound to a predefined role and execution order. These design choices make such systems effective within their target tasks like coding~\citep{hong2023metagpt,qian2023chatdev} and multimodal understanding~\citep{dai2025qoq,li2025puzzleworld,li2025unified,lin2025agent}, but difficult to generalize. \name\ differs on both axes: (1)~instead of a central executive~\citep{hong2023metagpt,qian2023chatdev,schmidgall2025agent}, \name\ uses up-tree competition and down-tree broadcast to determine information flow without a learned central orchestrator; (2)~instead of fixed workflow~\citep{hong2023metagpt,qian2023chatdev} or predefined tool-calling protocols~\citep{guo2024stabletoolbench,qin2023toolllm}, processors in \name\ have equal priorities is determined dynamically by competition at each iteration. \name\ also supports within-episode adaptation of processor memories and links with iterations, aligning with the recent progress of self-evolving agents~\citep{cemri2026adaevolve,novikov2025alphaevolve,qu2026coralautonomousmultiagentevolution,yang2025survey,zhou2025mem1}, but grounded in a cognitive architecture.

\section{\name: The Conscious Turing Machine with Modern AI Models}

In this section, we present background on the Conscious Turing Machine (CTM) (\S\ref{sec:ctm}), then describe how \name\
instantiates CTM's abstract architecture
(\S\ref{sec:components}) and dynamics
(\S\ref{sec:dynamics}) as a concrete and working system.

\subsection{Background on the Conscious Turing Machine}
\label{sec:ctm}

The CTM is a simple and formal model of consciousness~\citep{blum2022theory,blum2021theory} inspired by Alan Turing's model of computation~\citep{turing1936computable} and Bernard Baars' theater model of consciousness~\citep{baars1993cognitive}. However, CTM differs from Turing machines and Baars' model in several key ways. While Baars describes consciousness via the activity of actors performing on a stage directed by a stage director, the CTM has no stage director or central executive. Designing a central executive can be prohibitive since we often do not know how such an executive operates. Consider trying to recall the name of a person you've previously met. Although we may recall their name eventually, we do not know which processors are relevant and how to combine processor outputs beforehand. Rather, a federation of processors runs simultaneously, recalling different locations, events, and memories, before deciding which outputs are salient and integrating them to form the final answer. CTM employs a global workspace and distributed competition that determines which information from its vast collection of ``unconscious'' processors gets admitted to the ``conscious'' arena. When a problem needs to be solved, it becomes globally broadcast to all processors, eliciting help from those who might have the expertise, interest, and resources to tackle the problem, even though their talents and abilities might be unknown to a central executive. These features set the stage for its capability to be a model for general AI~\citep{blum2023agi}.

\subsection{CTM-AI Architecture}
\label{sec:components}
The formal definition of the CTM is a 7-tuple < STM, LTM, Up-Tree, Down-Tree, Links, Input, Output >. We provide a brief explanation for each of them here:
\begin{itemize}[noitemsep,topsep=0pt,nosep,leftmargin=*,parsep=0pt,partopsep=0pt]
    \item CTM is born at time $0$ and has a finite lifetime $T$, measured in discrete clock ticks, $t = 0, 1, 2, ..., T \approx 10^{10}$.
    \item $\textsc{STM}$ (short-term memory) is a small memory capable of holding a single chunk of information at each time $t$.
    \item $\textsc{LTM}$ (long-term memory) is a collection of $K$ powerful processors $\{p_1, p_2,..., p_K\}$, $K$ can be as large as $10^{7}$.
    \item Up-Tree is an up-directed binary tree of height $h$ with $K$ leaves, one leaf in each $\textsc{LTM}$ processor, and a (single) root in STM.
    \item Down-Tree is a simple down-directed tree of height $1$ with a single root in $\textsc{STM}$ and $K$ edges directed from that root to the leaves, one leaf in each $\textsc{LTM}$ processor.
    \item Links are the channels for transmitting information directly between processors.
    \item Input: $\mathbb{R}^d \rightarrow \textsc{LTM}$ carries information from the external (outer) world via sensors (\emph{e.g.}, eyes, ears) to special $\textsc{LTM}$ processors (\emph{e.g.}, visual and auditory processors). $\mathbb{R}^d$ is CTM’s external world where $\mathbb{R}$ represents the real numbers and $d$ is a positive integer. It also includes a user intent, like a query about the external world.
    \item Output: $\textsc{LTM} \rightarrow \mathbb{R}^d$ carries information from special processors (\emph{e.g.}, motor processor) that can be considered as feedback to the external (outer) world.
\end{itemize}


\begin{figure*}[t]
    \vspace{-0mm}
    \includegraphics[width=\textwidth]{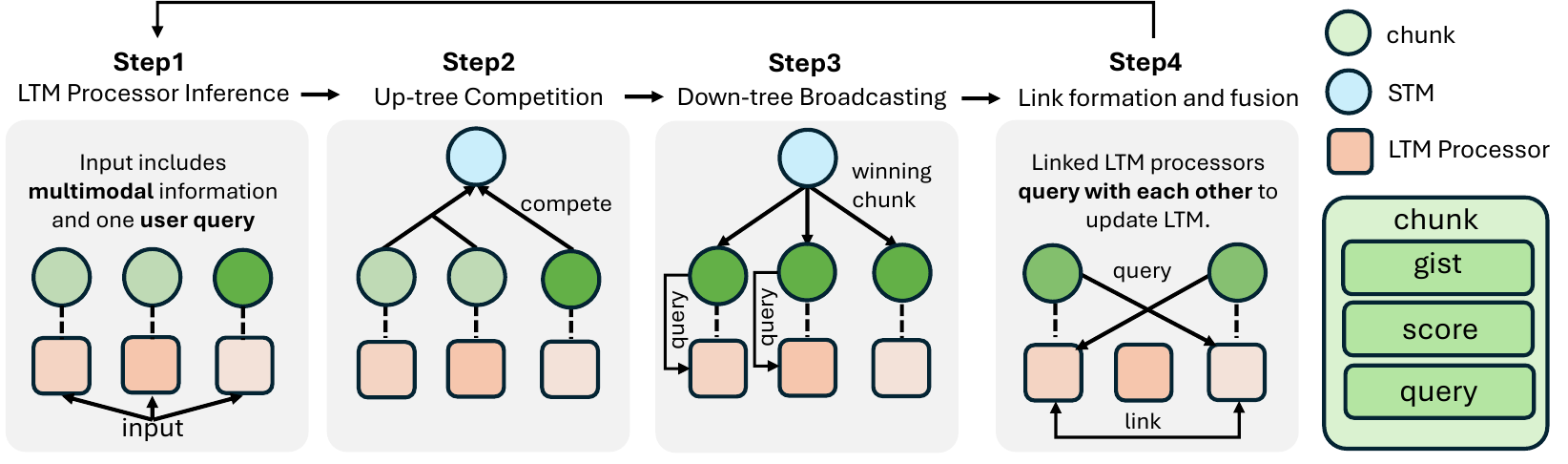}
    \caption{\small \textbf{Overview of \name dynamics}.
(\textbf{1})~all specialized \textsc{LTM} processors run in parallel,
each producing a chunk with a content gist, follow-up queries,
and a self-assessed score;
(\textbf{2})~an up-tree competition selects which chunk enters the
limited-capacity \textsc{STM}, determining the system's
conscious content;
(\textbf{3})~ A down-tree broadcast distributes this content to all processors, at which point the system becomes consciously aware of the selected information; (\textbf{4})~bidirectional links form between processors that hold
complementary information, enabling direct unconscious
communication that bypasses the \textsc{STM}, followed by fusion where linked processors exchange
to enrich their memories.
\name\ iterates with memories and links
evolving across iterations.}
    \vspace{-3mm}
    \label{fig:overview}
\end{figure*}

\xhdr{Long-term memory processors} The \name\ implementation comprises a configurable pool of LTM processors, each with its own expertise and memory. There are five broad families of LTM processors~\citep{card1980keystroke}:
\begin{itemize}[noitemsep,topsep=0pt,leftmargin=*,parsep=0pt,partopsep=0pt]
  \item Sensory processors convert raw perceptual signals (e.g., vision, language) into representations.
  \item Extended or artificial processors wrap external tools and APIs (e.g., calculators, web search, weather services) so that they can be accessed as internal modules. 
  \item Cognitive processors handle reasoning, inference, and planning for long-horizon problem solving. 
  \item Motor processors generate outputs by mapping internal intents to external actions, including dialogue utterances, API calls, or embodied movements. 
  \item Unspecialized ``free'' processors are expandable slots for future role discovery.
\end{itemize}

Formally, an LTM processor $p$ (with parameters $\theta$) operates in a shared space $\mathcal{H}$ and maintains a private memory state $M_t \in \mathcal{M}$ updated over time. Such a memory works as the context for the processor. At step $t$, it receives an observation $o_t\in\mathcal O$ and a user query $q_t\in\mathcal Q$. We view the LTM processor at time $t$ as a function $\mathrm{LTM}_t(\cdot)$ equipped with three operations: (1) \textit{execute} produces a chunk based on the current observations and previous memory; (2) \textit{read} returns a view of its memory at a specified timestamp; and (3) \textit{write} integrates one or more chunks into its memory:
{\small
\begin{equation}
    \textbf{execute:} \enspace {\mathrm{LTM}_t(o_t,q_t)=\mathbf{c}_t} \enspace \textbf{read:} \enspace \mathrm{LTM}_t(\cdot)=\mathrm{M}_t  \enspace 
\textbf{write:} \enspace \mathrm{LTM}_t(\mathbf{c}_t)= \mathrm{M}_t \oplus\mathbf{c}_t = \mathrm{LTM}_{t+1}(\cdot) 
\end{equation}
}

A chunk $\mathbf{c}_t$ produced by processor $p$ at step $t$ is defined as a tuple: $\mathbf{c}_t=\big\langle \mathrm{addr}(p),\, t,\, h_t,\, q_t,\, s_t \big\rangle$. Each chunk stores its unique identifier $\mathrm{addr}(p)$, the timestep $t$, a gist $h_t\in\mathcal{H}$ in English language that summarizes information relevant to the user's query (e.g., information from audio like laughter detected and likely humorous), one or more follow-up query $q_t\in\mathcal Q$ that the processor proposes to other processors if answering it could improve the final answer (e.g., a language processor can ask the vision processor for facial expressions), and a self-reported score $s_t$ indicating the processor’s confidence/utility for how useful the gist is to answer the query.

\xhdr{Short-term memory}
Short-term memory (STM) is a small memory that holds a single chunk of information at each time step $t$. After up-tree competition, the winning chunk $\textbf{c}_t^{i^\star}$ is stored in the STM. If its score $s_t^{i^\star}$ exceeds a threshold $\tau$, the chunk is considered as the conscious output of CTM-AI and sent as the output of the CTM-AI; otherwise, the STM would be broadcast to all LTM processors via down-tree broadcast, and the system proceeds to the next iteration to gather more information.

\subsection{\name\ Dynamics}
\label{sec:dynamics}

Based on this architecture, the following learning dynamics govern prediction and learning in CTM:
\begin{enumerate}[noitemsep,topsep=0pt,nosep,leftmargin=*,parsep=0pt,partopsep=0pt]
    \item Different $\textsc{LTM}$ processors perform distinct functions, \emph{e.g.}, cognitive, sensory, or motor. Some processors may be ``off-the-shelf'' while others' functionalities are realized over time. While individual processors may have their own internal language, communication within the CTM is in a common multimodal language we call Brainish. All processors start as independent entities.
    \item Conscious communication between processors is conducted via an Up-Tree competition that decides whose chunk of information gets into $\textsc{STM}$.
    \item The winning chunk (CTM's conscious content) is immediately globally broadcast to all processors via the Down-Tree, which causes the CTM to pay conscious attention to this information.
    \item Links between processors form over time as one processor views another as having relevant information, enabling unconscious communication to integrate their knowledge into higher-order information (\emph{e.g.}, learning to ride a bike requires conscious communication between sight and movement, after a while, links form, enabling unconscious communication).
    \item Through continuous interaction, feedback, and learning from its external world via sensory inputs, predictions, actuators, and feedback, the CTM updates its individual processors, processor links, and multiprocessor integration to improve over time.
\end{enumerate}

To implement these learning dynamics using modern AI architectures, we translate the first four CTM principles into a four-stage inference process for \name. The final principle dictates the overarching iteration loop. A formal description of each computational mechanism follows:

$\triangleright$ \xhdr{Step 1: LTM processor chunk inference}
At each iteration $t$, all $K$ LTM processors run in parallel on the observation $o_t$ and query $q_t$, each conditioned on its private memory $M_t^i$. Every processor jointly produces three outputs: a content gist $h_t^i$ summarizing its findings, a set of follow-up queries $q_t^i$ for potential cross-processor consultation, and a self-assessed score $s_t^i$. The resulting chunk $\mathbf{c}_t^i$ is:
\begin{equation}
    \mathrm{CTM}_{\mathrm{collect}}(o_t, q_t)
    = \left\{\mathrm{LTM}_t^i(o_t, q_t)\right\}_{i=1}^K
    = \{\mathbf{c}_t^i\}_{i=1}^K
    \label{eq:ctm_collect}
\end{equation}
We use $\operatorname{score}(\mathbf c_t^i)=s_t^i$
and $\operatorname{addr}(\mathbf c_t^i)=i$
to access the score and processor index stored in a chunk.

\textbf{Chunk score calculation}. Here we provide more information about the score function. Directly inspired by the CTM, the score $s_t^i$ is decomposed into three interpretable sub-scores: {weight} (how relevant the chunk addresses the query), {intensity} (the processor's confidence in its output), and {mood} (whether the chunk contains unexpected information) as an interpretable heuristic. These sub-scores are elicited alongside the chunk's gist via structured prompting. The final score is computed as a linear combination:
\begin{equation}
s_t^i
= \alpha_1 s_{\mathrm{weight},t}^i
+ \alpha_2 s_{\mathrm{intensity},t}^i
+ \alpha_3 s_{\mathrm{mood},t}^i
\label{eq:weight-decompose}
\vspace{2mm}
\end{equation}
where we set $\alpha_1=\alpha_2=1$ and $\alpha_3=0.2$,
down-weighting mood to prioritize reliable and on-topic chunks in the subsequent up-tree competition.

\label{sec:uptree}

$\triangleright$ \xhdr{Step 2: Up-tree competition into STM} After collecting chunks from all LTM processors, an up-tree competition selects the final chunk for STM's limited-capacity workspace. This competition is hierarchical and local: sibling chunks compete in groups via an additive competition function, ensuring the winning probability is independent of a processor's tree position. Theoretically, there can be a very large number of processors.
In practice, only a few ($< 10$) LTM processors are typically active during inference, so we adopt a simplified global competition that samples an STM entry from the chunk scores. Scores are normalized into a categorical distribution
(with temperature $\beta$). Here $\mathbf{s}_t=(s_t^1,\ldots,s_t^K)$
contains the self-reported scores already stored in
the collected chunks.
\begin{equation}
i^\star \sim \mathrm{Cat}\big(\mathrm{Softmax}(\mathbf{s}_t / \beta)\big), \quad
\mathrm{CTM}_{\mathrm{up}}(\{\mathbf{c}_t^i\}_{i=1}^K) = \mathbf{c}_t^{i^\star}.
\label{eq:ctm_up}
\end{equation}\label{sec:down-tree}
$\triangleright$ \xhdr{Step 3: Down-tree broadcast}
Once the up-tree competition selects the winning chunk
$\mathbf{c}_t^{i^\star}$, it is written into the STM. If the stopping criterion is not met, the winning chunk
is broadcast to all LTM processors via the down-tree.
The system becomes consciously aware of this
information upon reception by all processors~\citep{blum2022theory}. Operationally, each processor receives the broadcast and applies its own
\textbf{write} function to update its private memory:
\begin{equation}
\begin{aligned}
\mathrm{CTM}_{\mathrm{down}}(\mathbf{c}_t^{i^\star})
= \left\{\mathrm{LTM}_t^i(\mathbf{c}_t^{i^\star})\right\}_{i=1}^K
= \left\{\mathrm{LTM}_{t+1}^i(\cdot)\right\}_{i=1}^K
\end{aligned}
\label{eq:ctm_down}
\end{equation}
After this step, all processors share the same conscious content in their updated memories, preparing the system for
cross-processor integration in Step~4.

$\triangleright$ \xhdr{Step 4: Link formation and fusion}
While steps~1--3 represent \emph{conscious} communication in the CTM through STM broadcasts, step~4 enables \emph{unconscious} communication where LTM processors form links and exchange information with each other~\citep{blum2022theory}.

\textbf{Link formation}.
Triggered by the down-tree broadcast in Step~3, the winning processor $i^\star$ forms bidirectional
links with other processors based on their previously
reported chunk scores. Specifically, a link is established
between processors $i^\star$ and $j$ if
$\operatorname{score}(\mathbf c_t^j)>\eta$,
where $\eta$ is the link threshold.
The corresponding entries in the adjacency matrix are
updated as $L[i^\star,j],L[j,i^\star]\leftarrow 1$.
Since these scores are already generated during chunk
inference, link formation requires no additional
LLM calls or relevance evaluations.
For example, in sarcasm detection, vision, text, and
audio processors detect distinct cues (a sad face,
an angry tone, exaggerated speech) and form links
to share complementary evidence.
Links formed for a given datapoint remain permanent
across iterations.

\textbf{Link fusion}. Once links are established, each processor $\mathrm{LTM}_i$
consults its linked neighbors $\mathcal{N}(i)$ in parallel. It uses the follow-up queries $q_t^i$ generated
in Step~1 to consult its linked neighbors; the
neighbors respond via their \textbf{execute} function, and the initiating processor integrates these responses via
\textbf{write}:
{\setlength{\abovedisplayskip}{6pt}
\setlength{\belowdisplayskip}{0pt}
\begin{equation}
\begin{aligned}
\mathrm{CTM}_{\mathrm{fuse}}(o,\, L)
\!\!=\!\! \bigl\{\, \widehat{\mathrm{LTM}}_t^{i}
   \,\bigr\}_{i=1}^K
 \!\!=\!\! \left\{ \mathrm{LTM}_{t+1}^i(\cdot) \right\}_{i=1}^K, \ \ 
\widehat{\mathrm{LTM}}_t^{i}
\!=\! \mathrm{LTM}_t^{i}\!\Bigl(
   \bigl\{\mathrm{LTM}_t^{j}(o,\, q^{j})
   \bigr\}_{j \in \mathcal{N}(i)}\Bigr)
\end{aligned}
\label{eq:ctm_fuse}
\vspace{3mm}
\end{equation}
}This unconscious cross-processor integration discovers
richer, synergistic information that no single processor could
produce
alone~\citep{liang2024foundations,partan1999communication}.
The enriched long-term memories are carried into the next iteration of
Step~1, closing the inference loop.

\begin{wrapfigure}[16]{r}{0.52\textwidth}
\vspace{-22pt}
\begin{minipage}{0.52\textwidth}
\begin{algorithm}[H]
\footnotesize
\caption{\small CTM-AI Iterative Inference Algorithm}
\label{alg:ctm_ai}
\begin{algorithmic}[1]
\Require LTMs $\{\mathrm{LTM}^k\}_{k=1}^{K}$;
$\mathrm{STM}$; link matrix $L\in\{0,1\}^{K\times K}$;
query $q$; obs $o$; max rounds $R$;
thresholds $\tau,\eta$
\Ensure Output $\mathrm{STM}$ for answering query $q$

\For{$r = 1, \ldots, R$}

    \State $\{\mathbf c^i\}_{i=1}^K
    \gets \mathrm{CTM}_{\mathrm{collect}}(o,q)$ \Comment{Eq.~\eqref{eq:ctm_collect}}

    \State $\mathbf c^\star
    \gets \mathrm{CTM}_{\mathrm{up}}
    (\{\mathbf c^i\}_{i=1}^K)$
    \Comment{Eq.~\eqref{eq:ctm_up}}

    \State $\mathrm{STM}\gets\mathbf c^\star$

    \If{$\operatorname{score}(\mathbf c^\star)>\tau$
    }
        \State \Return $\mathrm{STM}$
    \EndIf

    \State $\{\mathrm{LTM}^i\}_{i=1}^K
    \gets \mathrm{CTM}_{\mathrm{down}}(\mathbf c^\star)$
    \Comment{Eq.~\eqref{eq:ctm_down}}

    \State $i^\star\gets\operatorname{addr}(\mathbf c^\star)$

    \State $L[i^\star,j],L[j,i^\star]\gets 1,\forall j\ne i^\star:
    \operatorname{score}(\mathbf c^j)>\eta$
    \State $\{\mathrm{LTM}^i\}_{i=1}^K
    \gets \mathrm{CTM}_{\mathrm{fuse}}(o,L)$
    \Comment{Eq.~\eqref{eq:ctm_fuse}}

\EndFor

\State \Return $\mathrm{STM}$

\end{algorithmic}
\end{algorithm}
\end{minipage}
\vspace{20pt}
\end{wrapfigure}

$\triangleright$ \xhdr{Overall: Iterative inference loop}
The CTM theory prescribes a continuous cycle of
\emph{prediction, feedback, and
learning}~\citep{blum2022theory}. CTM-AI keeps this
structure through Algorithm~\ref{alg:ctm_ai}.

\textbf{Prediction}.
All processors produce chunks from observations and memory, then compete via the up-tree to select STM content.

\textbf{Feedback}.
For agentic tasks (e.g., web navigation tasks), motor
processors translate the conscious content into actions on
the external environment. The environment's response returns
as a new observation $o_{t+1}$, providing feedback that informs the next step. For
non-agentic tasks (e.g., multimodal perception), no external
feedback is available, the system instead relies on iterative
internal refinement.

\textbf{Learning}.
In the original CTM, learning is realized through the
Sleeping Experts Algorithm, which adjusts processor weights
based on prediction outcomes. To achieve multi-agent evolution, CTM-AI instead leverages
in-context learning for self-reported score updates, requiring no parameter updates, through
two evolution mechanisms:
(1) \emph{memory evolution}: broadcast chunks and fused responses are written into each processor's private memory, enriching the context window for future inference; and
(2) \emph{structural evolution}: new links form
between processors with complementary information,
progressively densifying the communication graph for richer
information integration.

\section{Evaluating the Capabilities of \name}

We present quantitative results for \name\ on multimodal perception (\textsc{MUStARD}~\citep{castro2019towards} and \textsc{UR-Funny}~\citep{hasan2019ur}), tool use (\textsc{StableToolBench}~\citep{guo2024stabletoolbench}), and web agent tasks (\textsc{WebArena-Lite}~\citep{zhou2023webarena}).

\subsection{Evaluation Settings}

\xhdr{Unified model baselines}
We compare against strong unified single-model systems across all three evaluation axes. Here, ``unified'' means that a single model models multimodal interaction internally or manages multi-step tool use within one end-to-end system, without explicit decomposition into multiple collaborating agents.

For \textit{multimodal perception}, we include two types of baselines: ({1}) fine-tuned multimodal models, including MMoE~\citep{yu2023mmoe}, BLIP-2~\citep{li2023blip}, and ALBEF~\citep{li2021align}, which are trained to jointly model text and visual signals for multimodal interactions; and ({2}) prompting-based multimodal foundation models, including Qwen3-VL-8B-Instruct~\citep{bai2025qwen3}, Qwen3-VL-8B-Thinking~\citep{bai2025qwen3}, Qwen3-Omni-Flash~\citep{xu2025qwen3}, and Gemini-2.5-Flash-Lite~\citep{comanici2025gemini}.

For \textit{tool-use}, we compare with strong LLM-based agents, including GPT-4o~\citep{achiam2023gpt} and ToolLLaMA-v2~\citep{qin2023toolllm} under standard prompting strategies such as Chain-of-Thought and DFS-style planning. These baselines rely on a single LLM to plan and compose tool calls across multi-step tasks.

For \textit{web agent tasks}, we adopt ReAct~\citep{yao2023react} with GPT-4o and Gemini-2.5-Flash-Lite as base models. Both the ReAct baselines and the \name-based agent receive identical observations.

\xhdr{Multi-agent baselines}
To situate \name\ among existing multi-agent paradigms at comparable inference cost, we further compare with several representative frameworks: multi-agent debate~\citep{du2023improvingfactualityreasoninglanguage}, centralized orchestra~\citep{shen2023hugginggpt}, multi-agent ensembling~\citep{wang2022self}, MetaGPT~\citep{hong2023metagpt}, Mixture-of-Agents (MoA)~\citep{wang2024mixture}, and AutoGen~\citep{autogen_blog}. We instantiate all multi-agent baselines on the same backbone as \name\ for fair comparison, with each framework assigning dedicated agents to audio, video, text modalities, and tool use. Details are provided in Appendix~\S\ref{sec:exp_details}.

\begin{table}[t!]
  \centering
  \small
  \sisetup{
    detect-weight=true,
    table-number-alignment=center,
    mode = math,
    reset-text-family = false,
    reset-text-series = false,
    reset-text-shape = false,
    round-mode = places
  }

  \begin{minipage}[t]{0.48\linewidth}
    \centering
    \captionof{table}{\small \textbf{Evaluation results on multimodal perceptions (MUStARD and UR-FUNNY)}. We report accuracy and macro-F1. Fine-tuned models and Qwen3-VL-series take vision and text as inputs. Qwen3-Omni and Gemini-2.5-flash-lite take vision, text, and audio. Details in Appendix~\S\ref{sec:exp_details}.}
    \label{tab:mustard_UR-FUNNY}
    \resizebox{\linewidth}{!}{%
    \begin{tabular}{l
        S[table-format=2.2]
        S[table-format=2.2]
        S[table-format=2.2]
        S[table-format=2.2]
        }
        \toprule
        & \multicolumn{2}{c}{\textbf{MUStARD}} & \multicolumn{2}{c}{\textbf{UR-FUNNY}} \\
        \cmidrule(lr){2-3} \cmidrule(lr){4-5}
        \textbf{Method} & {{Acc}$\uparrow$} & {{F1}$\uparrow$} & {{Acc}$\uparrow$} & {{F1}$\uparrow$} \\
        \midrule
        \multicolumn{5}{l}{\textit{Fine-tuned models}}\\
        ALBEF & 54.49 & 48.51 & 66.77 & 68.67 \\
        BLIP2 & 53.75 & 62.65 & 70.43 & \cellcolor{bestgray} 74.31 \\
        MMoE  & 70.41 & 71.78 & 71.88 & 73.29 \\
        \midrule
        \multicolumn{5}{l}{\textit{Prompting-based models}}\\
        Qwen3-VL-8B-Instruct & 51.69 & 47.55 & 62.70 & 60.00 \\
        Qwen3-VL-flash       & 52.53 & 47.81 & 64.92 & 63.27 \\
        Qwen3-VL-8B-thinking & 57.58 & 55.61 & 67.84 & 67.60 \\
        \midrule
        Qwen3-Omni-flash     & 54.78 & 50.47 & 70.16 & 69.62 \\
        \rowcolor{ctmrowcolor} \ctmrow & 66.76 & 66.75 & 66.57 & 65.98 \\
        Gemini-2.5-fl & 71.91 & 71.71 & 70.70 & 70.77 \\
        \rowcolor{ctmrowcolor} \ctmrow & \cellcolor{bestgray} 72.28 & \cellcolor{bestgray} 72.23 & \cellcolor{bestgray} 72.13 & 72.13 \\
        \bottomrule
    \end{tabular}%
    }
  \end{minipage}
  \hfill
  \begin{minipage}[t]{0.48\linewidth}
    \centering
    \captionof{table}{\small \textbf{Evaluation results on tool-using (StableToolBench).} We report the solvable pass rate score evaluated with GPT-4o (MirrorAPI-Cache setting) . We focus on multi-tool calling scenarios. I2-Cat. stands for I2-Category, I2/I3-Inst. stands for I2/I3-Instruction. Details in Appendix~\S\ref{sec:exp_details}.}
    \label{tab:stabletoolbench}
    \resizebox{\linewidth}{!}{%
    \begin{tabular}{lccc}
        \toprule
        & \multicolumn{3}{c}{\textbf{StableToolBench}} \\
        \cmidrule(lr){2-4}
        \textbf{Method} & {I2-Cat.}$\uparrow$ & {I2-Inst.}$\uparrow$ & {I3-Inst.}$\uparrow$ \\
        \midrule
        ToolLLaMAv2 CoT & \valerr{19.9}{1.0} & \valerr{22.3}{0.4} & \valerr{19.1}{0.8} \\
        ToolLLaMAv2 DFS & \valerr{22.8}{1.5} & \valerr{19.2}{1.6} & \valerr{18.6}{1.5} \\
        GPT-4o mini CoT & \valerr{24.5}{1.0} & \valerr{22.3}{2.7} & \valerr{20.8}{1.5} \\
        GPT-4o mini DFS & \valerr{25.8}{1.7} & \valerr{25.8}{2.7} & \valerr{20.2}{0.8} \\
        GPT-4o CoT      & \valerr{32.5}{1.7} & \valerr{29.6}{1.6} & \valerr{27.9}{3.5} \\
        GPT-4o DFS      & \valerr{32.8}{1.5} & \valerr{28.3}{1.3} & \valerr{23.0}{1.3} \\
        \midrule
        Gemini-2.5-fl CoT  & \valerr{23.1}{2.1} & \valerr{12.7}{0.6} & \valerr{19.1}{0.8} \\
        \rowcolor{ctmrowcolor}\ctmrow & \valerr{47.4}{2.5} & \valerr{34.9}{1.9} & \valerr{46.7}{2.0} \\
        Qwen3-8B CoT       & \valerr{29.8}{2.4} & \valerr{26.2}{3.0} & \valerr{33.6}{1.2} \\
        \rowcolor{ctmrowcolor}\ctmrow & \valerr{52.1}{1.9} & \multicolumn{1}{c}{\cellcolor{bestgray}\valerr{53.4}{0.7}} & \multicolumn{1}{c}{\cellcolor{bestgray}\valerr{72.7}{1.0}} \\
        Qwen3-8B-think CoT & \valerr{39.7}{0.9} & \valerr{37.0}{0.7} & \valerr{39.1}{0.8} \\
        \rowcolor{ctmrowcolor}\ctmrow & \multicolumn{1}{c}{\cellcolor{bestgray}\valerr{56.5}{0.8}} & \valerr{43.1}{0.6} & \valerr{66.4}{0.7} \\
        \bottomrule
    \end{tabular}%
    }
  \end{minipage}
\vspace{-3mm}
\end{table}

\xhdr{CTM-AI backbone models}
For multimodal interaction, we instantiate processors in \name\ with Gemini-2.5-Flash-Lite and Qwen3-Omni-Flash, both of which natively accept text, audio, and vision inputs, and initialize each processor with the same underlying model so that comparisons against unified model baselines reflect architectural gains rather than differences in model capacity. For tool use, we additionally evaluate with Qwen3-8B-Instruct and Qwen3-8B-Thinking backbones under the same principle. We also use Gemini-2.5-Flash-Lite as the backbone model for agentic tasks. Details of per-task backbone assignments are provided in Appendix~\S\ref{sec:exp_details}.

\subsection{Evaluation Results}

\xhdr{\name\ consistently outperforms baselines across multimodal, tool-using, and agentic benchmarks}
CTM-AI consistently outperforms same-backbone and prompting-based baselines across all three evaluation axes. As shown in Table~\ref{tab:mustard_UR-FUNNY}, on multimodal perception tasks, we find that CTM-AI with Gemini-2.5-Flash-Lite achieves 72.28 F1 on MUStARD and 72.13 F1 on UR-FUNNY, matching or outperforming the strongest baselines. As shown in Table~\ref{tab:stabletoolbench}, on StableToolBench, we find that CTM-AI yields the most substantial gains, improving over the strongest single-model baseline (Qwen3-8B-thinking CoT) by up to +16.8 points on multi-tool scenarios. Similarly, as shown in Figure~\ref{fig:webarena_radar}, on WebArena-Lite, we find that CTM-AI improves over the ReAct baseline with the same backbone model across all five website categories. These results demonstrate that our proposed single decentralized multi-agent coordination mechanism effectively generalizes across distinct task families, from multimodal fusion in perception to multi-step planning in tool execution.

\xhdr{\name\ achieves better performance compared with unified cross-modal reasoning by coordinating multi-agents}
In our multimodal setting, each processor directly
observes only a single modality observation, rather than the
complete multimodal input. Nevertheless, CTM-AI
achieves competitive or even superior performance
to unified multimodal models through agentic
coordination, without additional cross-modal
joint training. Through competition, broadcast,
and fusion, specialized processors exchange
information to integrate complementary
evidence across modalities.
As shown in Table~\ref{tab:mustard_UR-FUNNY}
and Table~\ref{tab:stabletoolbench},
CTM-AI outperforms unified baselines under the same
Gemini-2.5-Flash-Lite backbone, achieving +0.5 F1
on MUStARD, +1.4 F1 on UR-FUNNY, and over +20 points
across all StableToolBench splits.
These improvements largely extend to Qwen3-Omni-Flash
and Qwen3-8B variants.
A notable exception is UR-FUNNY with Qwen3-Omni-Flash,
where positive-class predictions decrease from 64.0\%
to 36.9\%, improving precision (0.652 to 0.715)
but reducing recall (0.844 to 0.535) and macro-F1
(0.696 to 0.659).

\begin{table}[t]
\centering
\footnotesize
\setlength{\tabcolsep}{3pt}
\renewcommand{\arraystretch}{0.9}
\setlength{\aboverulesep}{0.3ex}
\setlength{\belowrulesep}{0.4ex}
\caption{\textbf{Comparison with multi-agent frameworks.}
We compare methods on macro-F1, wall-clock time per instance
($\Delta t$), and number of API calls per instance across
MUStARD and UR-FUNNY. All methods use Gemini-2.5-Flash-Lite
as the backbone. $R$ denotes the maximum number of
inference rounds. All agents inside share the same prompt
for inference. We further extend the same-backbone
comparison to StableToolBench and WebArena-Lite (lower panel),
reporting pass/success rate (\%), latency ($\Delta t$, s),
model calls (\#API), and external interactions
(\#Tool: tool calls; \#Act: environment actions),
under a separate evaluation setting from
Table~\ref{tab:stabletoolbench}.
Details are in Appendix~\S\ref{sec:exp_details}.}
\label{tab:efficiency_mustard_UR-FUNNY}

\sisetup{
  detect-weight=true,
  mode=math,
  reset-text-family=false,
  reset-text-series=false,
  reset-text-shape=false,
  table-alignment-mode=none,
  group-separator={,},
  group-minimum-digits=4,
}

\begin{tabular}{ll*{8}{S}}
\toprule
& Metric & {Unified} & {Ensemble} & {Orchestra}
& {Debate} & {MetaGPT} & {AutoGEN} & {MoA}
& {\name\ ($R{=}3$)} \\
\midrule

\multirow{3}{*}{\textbf{MUStARD}}
& F1$\uparrow$         & 71.7 & 69.3 & 55.6 & 63.8 & 66.1 & 71.1 & 72.0 & \cellcolor{gray!20}72.2 \\
& $\Delta t\downarrow$ & 7.2 & \cellcolor{gray!20}5.6 & 23.6 & 16.3 & 20.9 & 21.4 & 17.0 & 18.7 \\
& \#API$\downarrow$    & \cellcolor{gray!20}1.0 & 3.0 & 13.0 & 10.0 & 6.1 & 10.1 & 10.1 & 9.7 \\
\midrule

\multirow{3}{*}{\textbf{UR-FUNNY}}
& F1$\uparrow$         & 70.8 & 58.7 & 54.6 & 66.1 & 62.1 & 58.6 & 67.6 & \cellcolor{gray!20}72.1 \\
& $\Delta t\downarrow$ & 8.8 & \cellcolor{gray!20}5.4 & 24.2 & 17.9 & 27.0 & 37.1 & 36.7 & 42.8 \\
& \#API$\downarrow$    & \cellcolor{gray!20}1.0 & 3.0 & 13.0 & 10.0 & 6.1 & 9.9 & 10.2 & 17.0 \\

\midrule[\heavyrulewidth]
& Metric & {ReAct} & {Ensemble} & {Orchestra}
& {Debate} & {MetaGPT} & {AutoGEN} & {MoA}
& {\name\ ($R{=}3$)} \\
\midrule

\multirow{4}{*}{\textbf{StableToolBench}}
& Pass$\uparrow$       & 32.9 & 38.1 & 56.5 & 38.0 & 46.3 & 44.0 & 58.3 & \cellcolor{gray!20}62.9 \\
& $\Delta t\downarrow$ & \cellcolor{gray!20}85 & 160 & 171 & 316 & 102 & 98.9 & 356 & 186 \\
& \#API$\downarrow$    & \cellcolor{gray!20}5.2 & 16.5 & 14.7 & 32.4 & 6.4 & 9.4 & 19.4 & 12.5 \\
& \#Tool$\downarrow$   & \cellcolor{gray!20}3.6 & 10.8 & 7.6 & 22.0 & 3.9 & 5.6 & 10.7 & 10.6 \\
\midrule

\multirow{4}{*}{\textbf{WebArena-Lite}}
& SR$\uparrow$         & 12.1 & 20.7 & 16.4 & 17.9 & 16.4 & 17.9 & 20.0 & \cellcolor{gray!20}25.1 \\
& $\Delta t\downarrow$ & 122 & 118 & 128 & 152 & 112 & 145 & 130 & \cellcolor{gray!20}106 \\
& \#API$\downarrow$    & \cellcolor{gray!20}7.7 & 28.6 & 38.4 & 49.4 & 27.4 & 47.9 & 31.3 & 38.2 \\
& \#Act$\downarrow$    & 7.7 & 7.2 & 7.7 & 7.0 & 9.2 & 6.8 & 7.8 & \cellcolor{gray!20}5.2 \\

\bottomrule
\end{tabular}
\vspace{-5mm}
\end{table}

\xhdr{\name\ outperforms other multi-agent frameworks with adaptive cost trade-offs}
CTM-AI overcomes the typical accuracy-flexibility trade-off of modular multi-agent pipelines, outperforming six representative multi-agent frameworks under the same backbone. As shown in Table~\ref{tab:efficiency_mustard_UR-FUNNY}, we find that CTM-AI achieves the highest F1 on both MUStARD (72.23) and UR-FUNNY (72.13), outperforming the strongest multi-agent baseline (MoA) by +0.2 and +4.6 F1, respectively. With only two inference rounds ($R=2$), CTM-AI achieves F1 scores of 69.8 and 70.0 on MUStARD and UR-FUNNY, respectively, using 6.9 and 9.4 API calls per instance.
On StableToolBench and WebArena-Lite, CTM-AI ($R=3$)
further achieves a 62.9\% pass rate and a 25.1\%
success rate, outperforming the strongest baselines
by 4.6 and 4.4 percentage points, respectively.
These gains also come at competitive inference costs:
CTM-AI uses fewer API calls than MoA on StableToolBench
(12.5 vs.\ 19.4) and achieves the lowest latency on
WebArena-Lite (106 seconds).
These results highlight two key insights.
First, CTM-AI's dynamic coordination is particularly
beneficial when contextual cues are sparse, as in
UR-FUNNY (+4.6 F1 over MoA versus +0.2 on MUStARD).
Second, iterative inference enables an adaptive
accuracy--cost trade-off, allowing easier instances
to stop early while harder ones receive additional refinement.

\subsection{Discussions}

\begin{wraptable}{r}{0.44\textwidth}
  \vspace{-4mm}
  \centering
  \footnotesize
  \sisetup{
    detect-weight=true,
    table-number-alignment=center,
    mode=math,
    reset-text-family=false,
    reset-text-series=false,
    reset-text-shape=false,
    round-mode=places
  }
  \setlength{\tabcolsep}{3pt}          
  \renewcommand{\arraystretch}{0.9}    
  \setlength{\aboverulesep}{0.2ex}     
  \setlength{\belowrulesep}{0.3ex}     

  \caption{\small \textbf{Scalability of processor number $K$}.
  We report the solvable pass rate on StableToolBench (I2-Cat., I2-Inst., I3-Inst.),
  model calls (\#API), latency ($\Delta t$, s), and average active links (\#Links) for different $K$.}
  \label{tab:k_ablation}
  \vspace{-1mm}

  \begin{tabular}{c
      S[table-format=2.1, round-precision=1]
      S[table-format=2.1, round-precision=1]
      S[table-format=2.1, round-precision=1]
      S[table-format=3.1, round-precision=1]
      S[table-format=3.0, round-precision=0]
      S[table-format=1.2, round-precision=2]}
      \toprule
      & \multicolumn{3}{c}{Pass rate$\uparrow$} & & & \\
      \cmidrule(lr){2-4}
      \textbf{$K$} & {I2-Cat.} & {I2-Inst.} & {I3-Inst.} & {\#API} & {$\Delta t$} & {\#Links} \\
      \midrule
       1 & 35.1 & 34.2 & 56.2 & 5.6   & 67  & 0.00 \\
       2 & 46.6 & 44.4 & 46.3 & 10.5  & 111 & 0.20 \\
       4 & 45.8 & 48.7 & 70.2 & 18.1  & 141 & 0.54 \\
       8 & 51.6 & 53.2 & 63.9 & 31.7  & 168 & 0.89 \\
      16 & \cellcolor{gray!20}61.2 & \cellcolor{gray!20}54.0 & 68.6 & 58.2 & 285 & 1.58 \\
      32 & 55.7 & 53.4 & \cellcolor{gray!20}77.9 & 108.0 & 427 & 2.33 \\
      \bottomrule
  \end{tabular}
  \vspace{-4mm}
\end{wraptable}

\vspace{-1mm}
\textbf{Q1: Can CTM-AI keep improving its performance when the number of processors keeps increasing?}

We vary the processor pool from $K$=1 to $K$=32 on StableToolBench while keeping everything else fixed. The macro-average of the pass rate rises steadily from 41.8 to 62.3, but with diminishing returns: the largest gain comes from $K$=2 to 4, while 16 to 32 adds only 1.0.
Meanwhile, model calls nearly double with each doubling of $K$ (5.6 $\rightarrow$ 108.0), latency grows from 67 to 427, and active links increase from 0 to 2.3.
Notably, $K$=16 already reaches 98\% of the best score with about half the model calls of $K$=32, offering a favorable accuracy--cost trade-off.

\textbf{{Q2}: Without external feedback, why can self-reported scores in CTM-AI help self-improve?}

\begin{figure*}[t]
    \centering
    \begin{minipage}[t]{0.32\textwidth}
        \centering
        \includegraphics[width=0.85\linewidth]{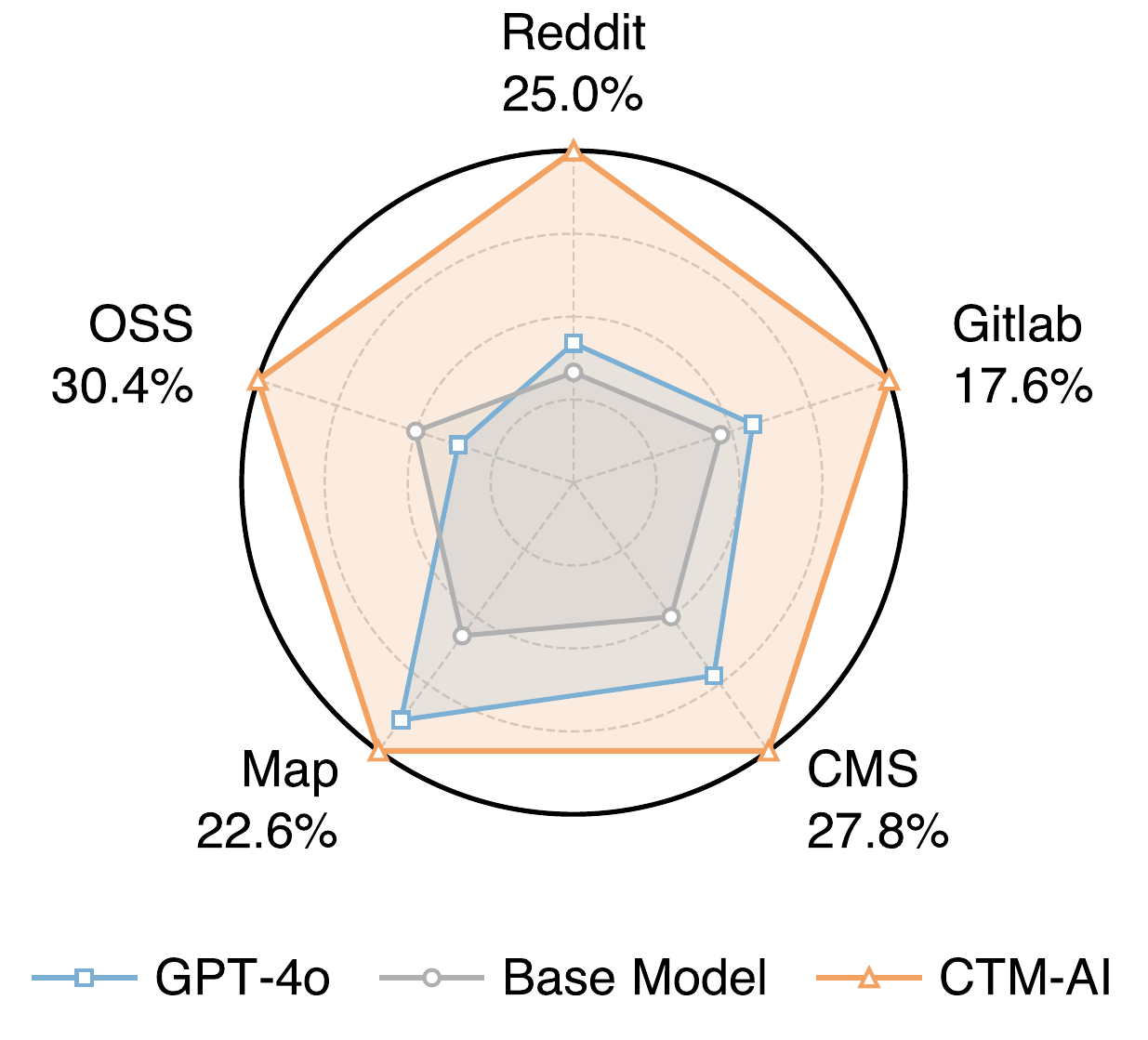}
        \caption{\small \textbf{Evaluation results on agentic tasks (\textsc{WebArena-Lite})}. Base model represents ReAct-style Gemini-2.5-flash-lite and \name\ uses the same backbone model. We report the success rate across 5 sub-domains in web agent tasks.}
        \label{fig:webarena_radar}
    \end{minipage}
    \hfill
    \begin{minipage}[t]{0.32\textwidth}
        \centering
        \includegraphics[width=\linewidth]{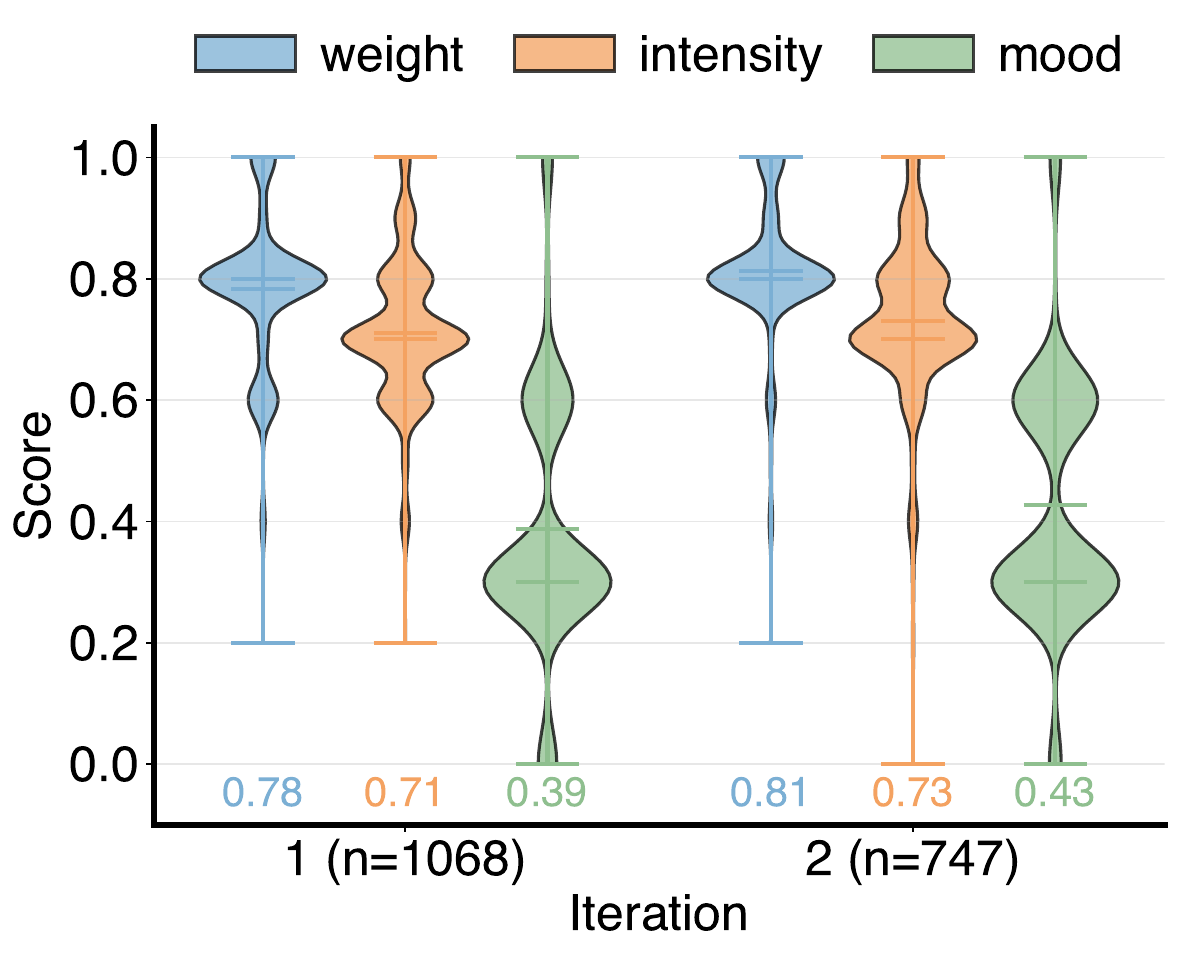}
        \vspace{-4mm}
        \caption{\small \textbf{Distribution of self-reported scores.} We summarize the score distribution of weight, intensity, and mood in each iteration, using Gemini-2.5-Flash-Lite as the backbone on the MUStARD dataset. $n$ is the processor number.}
        \label{fig:score_distributions}
    \end{minipage}
    \vspace{-2mm}
    \hfill
    \begin{minipage}[t]{0.32\textwidth}
        \centering
        \includegraphics[width=\linewidth]{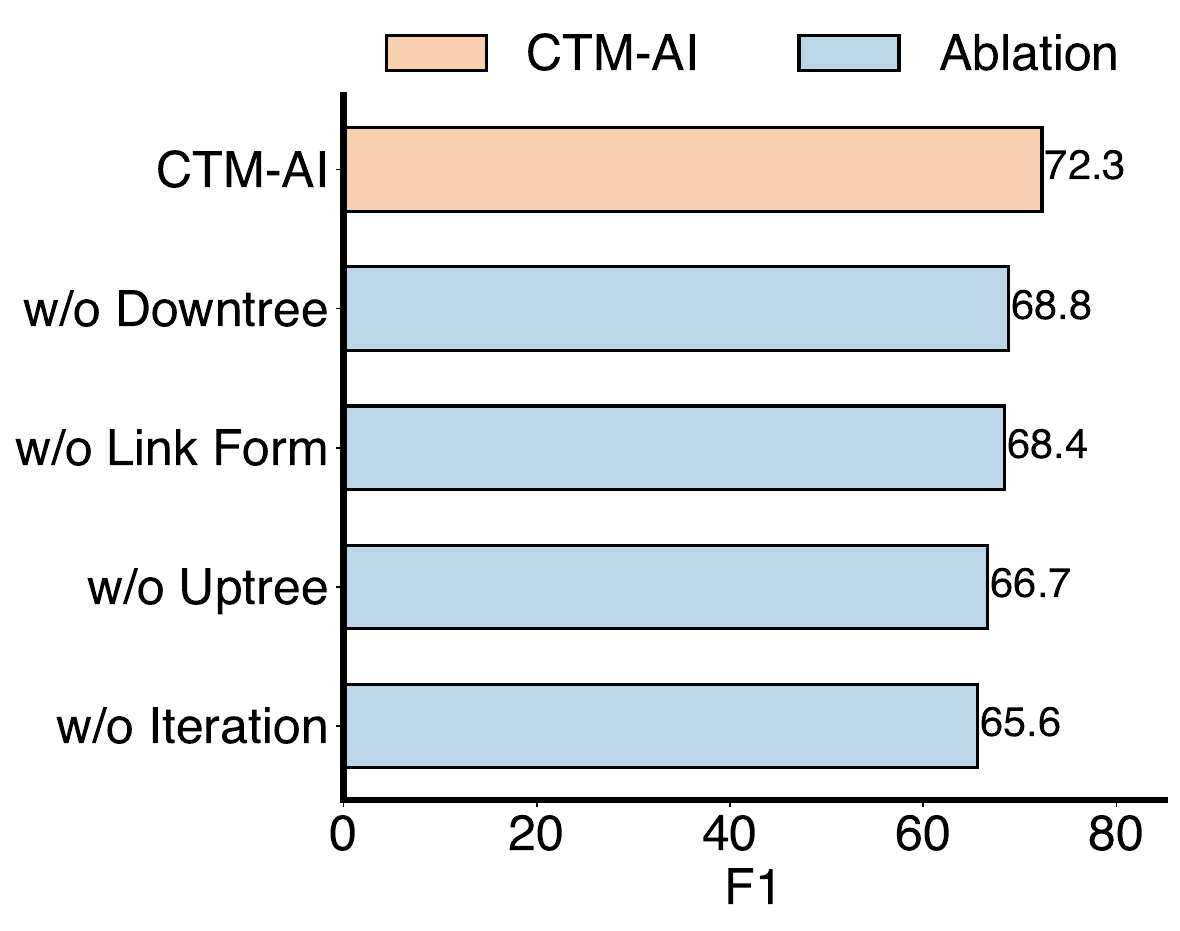}
        \vspace{-4mm}
        \caption{\small \textbf{Ablation on CTM-AI dynamics.} We isolate the contribution 
of each CTM-AI mechanism by ablating Step~2-4 and the iterative loop individually, using 
Gemini-2.5-Flash-Lite as the backbone of the MUStARD dataset. }
        \label{fig:ablation_components}
    \end{minipage}
    \vspace{-2mm}
\end{figure*}

\textbf{Multi-dimensional score decomposition provides an effective routing signal.} Rather than asking each processor for a single scalar score, we decompose chunk scoring into three interpretable sub-scores—weight, intensity, and mood (Eq.~\ref{eq:weight-decompose})—to yield a discriminative ranking signal for up-tree competition. Figure~\ref{fig:score_distributions} shows that these sub-scores exhibit well-differentiated distributions, with both weight and intensity spanning 0.6 to 1.0. In addition, our ablation study in Appendix~\S\ref{sec:self_reported_score_ablation} demonstrates that replacing CTM-AI's weighted sampling with minimum- or random-selection drastically degrades performance: on MUStARD, F1 drops from 72.23 to 62.92 ($-9.3$) under min-selection and to 66.76 ($-5.5$) under random-selection; on UR-FUNNY, it drops from 72.13 to 53.07 ($-19.1$) and 66.07 ($-6.1$), respectively. Our structured score also achieves a 62.9\% downstream pass rate, outperforming self-prompted Yes/No logits (55.0\%). Together, these results highlight two key insights: (1) pairing query alignment (weight) with a down-weighted tiebreaker (mood) reliably filters out low-quality processor chunks; and (2) effective multi-agent routing hinges on relative ranking utility rather than strict probabilistic calibration.

\textbf{Iterative multi-agent evolution progressively refines processor outputs until convergence.} Through global broadcast and link fusion, processors leverage cross-processor context across rounds to successively improve their predictions. Tracking the distribution of self-reported sub-scores across iterations on MUStARD (Figure~\ref{fig:score_distributions}) reveals that from iteration 1 to 2, weight ($0.78 \to 0.81$), intensity ($0.71 \to 0.73$), and mood ($0.39 \to 0.43$) all increase before stabilizing by iteration 3. This progression yields a key insight: even without external environmental feedback, global broadcast acts as an internal calibration signal, enabling processors to refine their generations, build confidence across rounds, and naturally converge to a stronger final output.

\textbf{{Q3}: As an inference-time scaling method, how does CTM-AI compare with reasoning?}

\textbf{CTM-AI and reasoning scale along orthogonal, stackable axes.}
While standalone reasoning models improve over base models by $+10$ points on StableToolBench and $+5$ points on MUStARD/UR-FUNNY (Tables~\ref{tab:mustard_UR-FUNNY} and \ref{tab:stabletoolbench}), they still underperform CTM-AI under a fixed backbone because chain-of-thought alone cannot coordinate across modalities and tools. Crucially, these two scaling paradigms are complementary: applying CTM-AI on top of Qwen3-8B-think yields an additional $+16.8$ to $+27.3$ points on StableToolBench over Qwen3-8B-think alone. Mechanistically, whereas internal reasoning deepens deliberation \emph{within} each processor, CTM-AI scales coordination \emph{across} processors. Stronger local reasoning produces higher-quality chunks and sharper queries throughout the cycle, causing the two gains to compound rather than overlap.

\textbf{Upgrading a single processor with a reasoning model elevates the entire system.}
Upgrading only the text processor from Gemini-2.5-Flash-Lite to OpenAI o3 boosts overall MUStARD F1 from $72.23$ to $78.11$ ($+5.88$), surpassing standalone o3. Notably, the stronger text processor does not dominate the workspace: winning chunks remain balanced across audio ($52.2\%$), video ($25.3\%$), and text ($22.5\%$), yet accuracy on text-winning samples surges to $82.46\%$.
This result is because modularity allows stronger reasoning to be injected into a single modality without retraining the rest, uplifting peer processors rather than overriding them. Specifically, o3 not only calibrates its own chunk scores more selectively (hence the high precision at a $22.5\%$ win rate), but also replaces generic broadcasts (e.g., \emph{``What is the comedic tone?''}) with targeted cross-modal queries (e.g., \emph{``Does Chandler roll his eyes?''}), guiding the other processors in the CTM-AI, including audio and video ones, to extract more discriminative features (Appendix~\S\ref{appendix:additional-case-study}).

\begin{figure*}[t]
    \centering
    \begin{minipage}[t]{0.31\textwidth}
        \centering
        \includegraphics[width=\linewidth]{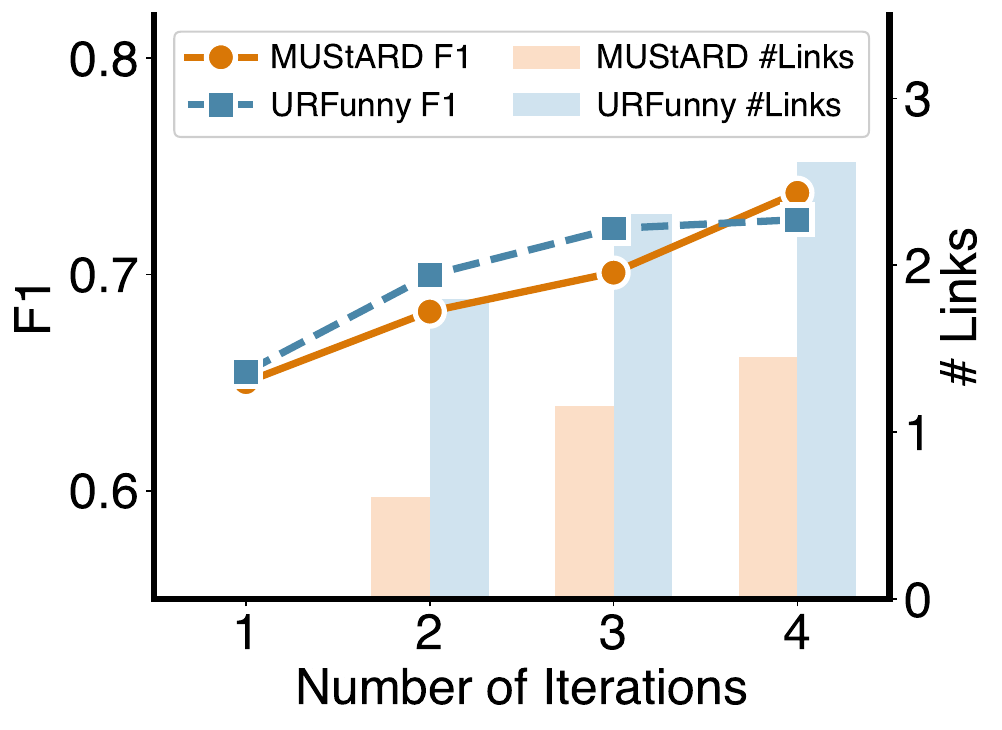}
        \vspace{-4mm}
        \caption{\small \textbf{Ablation on max iteration number $T$}. We use $\tau=2.2$, $\eta=0.9$ for MUStARD and $\tau=2.2$, $\eta=0.7$ for UR-FUNNY. When $T=1$, no links are formed.}
        \label{fig:ablation_iteration}
    \end{minipage}
    \hfill
    \begin{minipage}[t]{0.31\textwidth}
        \centering
        \includegraphics[width=\linewidth]{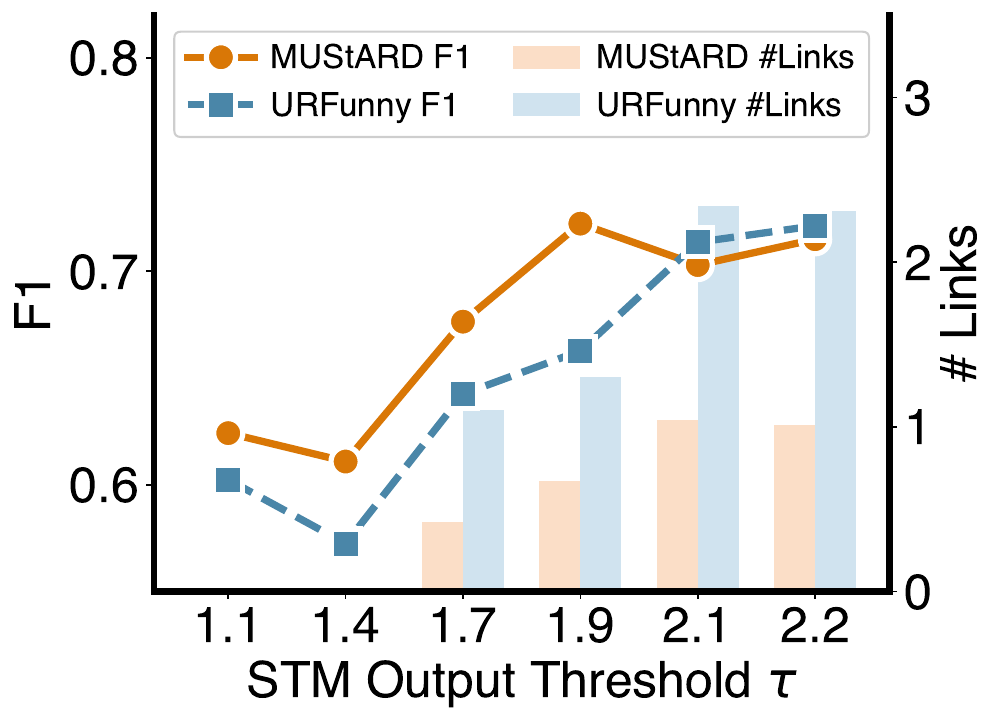}
        \vspace{-4mm}
        \caption{\small \textbf{Ablation on STM output threshold $\tau$.} We keep $\eta=0.9$ for MUStARD and $\eta=0.7$ for UR-FUNNY, with 3 max iterations for MUStARD and UR-FUNNY.}
        \label{fig:ablation_stm_threshold}
    \end{minipage}
    \hfill
    \begin{minipage}[t]{0.31\textwidth}
        \centering
        \includegraphics[width=\linewidth]{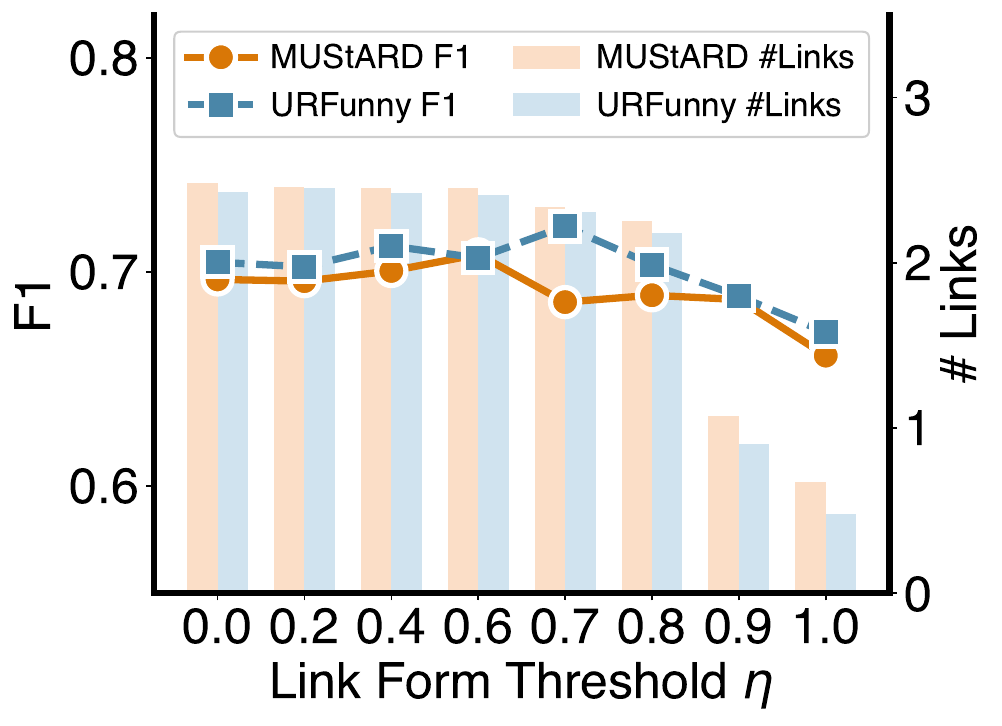}
        \vspace{-4mm}
        \caption{\small \textbf{Ablation on link form threshold $\eta$.} The higher $\eta$, the harder to form links.  We set $\tau=2.2$ and 3 max iterations for both MUStARD and UR-FUNNY.}
        \label{fig:ablation_link_form}
    \end{minipage}
    \vspace{-3mm}
\end{figure*}

\subsection{Ablation Studies}

\xhdr{Ablation on \name\ dynamics}
\name\ relies on five mechanisms: (i) processor inference, (ii) up-tree competition, (iii) down-tree broadcast, (iv) link formation and fusion, and (v) the iterative loop. Processor inference is always required, so we disable each of the other four in turn.
As shown in Figure~\ref{fig:ablation_components}, removing any single mechanism degrades performance. Ranked by impact, the losses are: iterative loop (-6.7 F1), up-tree competition (-5.6), link fusion (-3.9), and down-tree broadcast (-3.5).
Such results indicate that each mechanism has a distinct role, which explains the ranking. The iterative loop decides whether the system can refine its output at all. Up-tree competition makes sure the most informative chunk wins attention. Fusion integrates information across processors. Broadcast keeps all processors in sync. Even the smallest drop is substantial (-3.5 F1), so every mechanism is needed and \name\ is not over-engineered.

\xhdr{Ablation on max iteration number $T$}
As shown in Figure~\ref{fig:ablation_iteration}, increasing $T$ from 1 to 4 improves performance steadily on both datasets. MUStARD rises from 65.0 to 73.8 (+8.8), and UR-FUNNY rises from 65.5 to 72.5 (+7.0). The average number of inter-processor links grows along with performance. At $T{=}4$, UR-FUNNY forms 2.6 links on average, compared with 1.5 for MUStARD.
These results indicate that extra iterations pay off primarily by enabling denser link formation. Notably, while performance trends are similar across both datasets, their link counts differ—demonstrating that \name\ adaptively scales its cross-processor interaction to task demands without manual tuning.

\xhdr{Ablation on thresholds $\tau$ and $\eta$}
We test robustness to two hyperparameters. The first is the short-term memory (STM) output threshold $\tau$: a higher $\tau$ means stricter output filtering. The second is the link formation threshold $\eta$: a higher $\eta$ makes it harder for two processors to form a link.
As shown in Figure~\ref{fig:ablation_stm_threshold}, raising $\tau$ generally improves performance on both MUStARD and UR-FUNNY, together with more inter-processor links. As shown in Figure~\ref{fig:ablation_link_form}, performance is best for $\eta \in [0.4, 0.7]$ on both datasets. Both performance and the number of links drop sharply once $\eta > 0.8$.
Such results indicate that a stricter $\tau$ stops the system from committing to an answer too early. This leaves more rounds for processors to form links and exchange evidence. $\eta$ shows a clear trade-off: if it is too low, weak links add noise, and if it is too high, useful cross-processor fusion is cut off. The same good ranges hold on both datasets, so \name\ is reasonably robust to these hyperparameters.

\section{Conclusion}

In this paper, we introduced \name, a self-improving multi-agent framework that translates the theoretical principles of the Conscious Turing Machine (CTM) into a practical AI system. Through strong performance across multimodal perception, tool use, and agentic tasks, \name\ establishes a robust foundation for cognitively inspired multi-agent collaboration and iterative refinement. Moving beyond philosophical debates on machine consciousness, we hope our work inspires future research toward building more capable, decentralized, and self-improving multi-agent architectures.




{
\small
\bibliographystyle{plainnat}
\bibliography{example_paper}
}

\clearpage
\appendix
\newtheorem{prompt}{Prompt}
\appendix
\onecolumn

\section{Impact Statement}
\label{impact_statement}
This work utilizes publicly available datasets; no private or sensitive user data was collected, and all experiments were conducted in controlled research settings. By bridging the theoretical CTM framework with practical AI technologies, we aim to enhance AI capabilities in affective learning, decision-making, multi-step reasoning, and tool use, contributing to the development of more reliable and trustworthy general AI. Crucially, our goal is neither to build conscious AI nor to replicate human identity, thereby avoiding the ethical risks associated with deceptive anthropomorphization. We also acknowledge the inherent risks of deploying foundation models and AI agents, particularly the propagation of socio-cultural biases. Actively detecting, understanding, and mitigating these biases remains a central commitment of our ethical research framework.

\section{Limitations}
\label{sec:limitations}
\textbf{Scale of processor federation.} 
The original CTM theory envisions a federation of up to $10^7$ LTM processors. Our principal tasks typically use fewer than ten active processors (three modalities or roughly six retrieved tool processors); a separate $K=1$--$32$ study and a 512-processor stress test do not approach the theoretical scale. As a result, the simplified global competition we adopt in Step 2 (Eq.~\ref{eq:ctm_up}) approximates—but does not fully exercise—the hierarchical up-tree competition prescribed by the theory. Scaling CTM-AI to substantially larger and more heterogeneous processor pools, and studying whether hierarchical competition yields additional benefits at scale, remains an open direction. The 512-processor stress test requires approximately 1,022 model calls and 931k tokens for a query and is evidence of operational feasibility only, not scalable efficiency.

\textbf{Inference cost.}
Although CTM-AI achieves competitive latency relative to other multi-agent frameworks (Table~\ref{tab:efficiency_mustard_UR-FUNNY}), its iterative inference loop and link-fusion stage still incur higher API cost than unified single-model baselines. On UR-FUNNY, CTM-AI$_3$ uses $17.0$ API calls per instance versus $1.0$ for the unified baseline. Reducing this overhead is a natural next step.

\section{The Use of Large Language Models (LLMs)}
\label{ai-assistant-details}

We used LLMs as a writing assistant to help us edit parts of the paper. Additionally, we utilize the power of CodePilot and Claude Code to help us code faster. All AI-generated writing and code are manually checked and modified. There is no fully AI-generated content in the paper.

\section{Artifact Details}
\label{appendix:artifact-details}

\subsection{Model License}

\textbf{GPT-4o}~\citep{openai2024gpt4o} License: Proprietary (OpenAI)\\
\textbf{OpenAI-o3}~\citep{openai2025o3} License: Proprietary (OpenAI)\\
\textbf{Gemini-2.5-flash-lite}~\citep{comanici2025gemini} License: Apache 2.0\\
\textbf{Qwen3-VL-8B-Instruct}~\citep{bai2025qwen3} License: Apache 2.0\\
\textbf{Qwen3-VL-8B-thinking}~\citep{bai2025qwen3} License: Apache 2.0\\
\textbf{Qwen3-Omni-flash}~\citep{xu2025qwen3} License: Proprietary (Alibaba)

\subsection{Software Versions}
For web-agent evaluation, we adopt BrowserGym v0.14.2~\footnote{{\url{https://github.com/ServiceNow/BrowserGym}}}. To access large language models, we employ LiteLLM 1.74.3~\footnote{{\url{https://litellm.ai}}} as the serving interface.

\subsection{Dataset Statistics}
We include the test sets of MUStARD, URFunny, WebArena-Lite, and StableToolBench for evaluation. Table~\ref{tab:dataset_stats} shows their statistics.

\textbf{MUStARD}~\citep{castro2019towards} is a multimodal sarcasm detection dataset collected from TV shows, where each instance consists of an utterance with its conversational context and a binary sarcasm label. We evaluate on its 356-instance test split.

\textbf{UR-Funny}~\citep{hasan2019ur} is a multimodal humor detection dataset built from TED talks, where the task is to predict whether a punchline is humorous given its textual, visual, and acoustic context. We evaluate on its 992-instance test split.

\textbf{WebArena-Lite}~\citep{liu2024visualagentbench} is a human-verified subset of WebArena~\citep{zhou2023webarena} comprising tasks across five self-hosted websites: Reddit, GitLab, CMS, Map, and OneStopShop (OSS). Each task requires an agent to complete a natural-language instruction by interacting with real web pages, and is evaluated by programmatic success checkers. 

\textbf{StableToolBench}~\citep{guo2024stabletoolbench} extends ToolBench~\citep{qin2023toolllm} with a stabilized virtual API server and a solvability-filtered query set for tool-use evaluation. We evaluate on the I2-Inst, I2-Cat, and I3-Inst subsets using MirrorAPI-Cache~\citep{guo2025stabletoolbench} as the tool environment, which fine-tunes a specialized LLM on StableToolBench's cached API traces to stably mirror real API behaviors. As noted by~\citep{guo2025stabletoolbench}, some queries in the original test set reference APIs that are no longer available; such queries may fail during evaluation regardless of the agent's behavior. We report results over queries with valid APIs in each subset.

\section{Experimental Details}
\label{sec:exp_details}
In this section, we provide more implementation details related to the algorithm that we proposed based on \name\. We also include the prompting details to explain how we adapt \name\ architecture to different types of tasks.

\subsection{Details of Backbone Models}
We select Gemini-2.5-flash-lite as our base model to make most of the processors. It is mainly because Gemini-2.5-flash-lite is relatively small-scale and supports audio, vision, and text as input for inference. When querying the Gemini API, we adopt a deterministic decoding configuration with temperature fixed at 0.1, top-n set to 1, and a maximum token limit of 4096.

\subsection{Details of Unified Model Baselines}
We evaluate unified model architectures across four distinct tasks, encompassing both fine-tuned models and prompting-based foundation models.

\xhdr{Sarcasm Detection (MUStARD)} For fine-tuned baselines, we evaluate ALBEF~\citep{li2021align} (209.5M parameters), BLIP2~\citep{li2023blip} (2.7B parameters), MMoE~\citep{yu2023mmoe}, and our BaseModel (Gemini-2.5-flash-lite). For prompting-based baselines, we evaluate Qwen3-VL-8B-Instruct, Qwen3-VL-8B-thinking, Qwen3-Omni-flash, and Gemini-2.5-flash-lite using identical zero-shot/few-shot prompts.

\xhdr{Humor Detection (URFUNNY)} To assess multimodal affective understanding, we evaluate this task using the same comprehensive suite of unified models as the MUStARD task, including both the fine-tuned architectures (ALBEF, BLIP2, MMoE, Gemini BaseModel) and prompting-based foundation models (Qwen3 variants, Gemini-2.5-flash-lite).

\xhdr{API Tool Calling (StableToolBench)} We evaluate ToolLLaMA v2~\citep{qin2023toolllm} as the fine-tuned baseline, which is explicitly trained on the benchmark's train set. For prompting and search-based models, we compare GPT-4o-mini, GPT-4o, Qwen3-8B, Qwen2-8B-think, and Gemini-2.5-flash-lite, utilizing Chain-of-Thought (CoT)~\citep{wei2022chain} and Depth-First Search (DFS) strategies via MirrorAPI-Cache.

\xhdr{Web Navigation (WebArena-Lite)} We evaluate standalone foundation models in a direct agentic setting without multi-agent orchestration, specifically comparing the unassisted generation capabilities of GPT-4o and Gemini-2.5-flash-lite. We all use React as an agentic strategy that interleaves reasoning and acting to solve complex tasks. We use this as the primary baseline framework for the WebArena-Lite environment, driven by GPT-4o and Gemini-2.5-flash-lite backbones.

\subsection{Details of Multi-Agent Baselines}
To assess the advantages of \name's dynamics against existing multi-agent paradigms, we compare it against the following frameworks. Unless otherwise specified, these baselines utilize Gemini-2.5-flash-lite as the backbone model and maintain consistent hyperparameters ($T=0.2$, layers $N=3$).

\xhdr{Multi-Agent ensemble (Ensemble)} We conduct inference for all processors in the CTM-AI (video, audio, and text processor). Each processor outputs one answer, and we directly conduct majority voting on all of them to make the final answer.

\xhdr{Multi-agent debate (Debate)} A collaborative framework where multiple agents argue from different viewpoints over successive rounds. At each round, every agent observes the previous round's responses from all other agents and is asked to either defend its position or revise its answer; a separately prompted judge model then aggregates the final round into a single decision. We instantiate three debaters (one per modality: video, audio, text) plus one judge, and strictly cap the total debate depth at 10 API calls per example.

\xhdr{Multi-agent centralized orchestra (Orchestra)} A structured three-stage workflow inspired by planner--executor--aggregator designs. A \emph{controller} first decomposes the input task into a set of self-contained sub-queries; an \emph{executor} pool answers each sub-query independently and in parallel; finally, a \emph{summarizer} aggregates the sub-answers and produces the final prediction. Only the executor stage is parallelizable; the controller and summarizer stages run sequentially.

\xhdr{Mixture-of-Agents (MoA)}~\citep{wang2024mixture} A layered multi-agent architecture with $N=3$ sequential layers, each consisting of one video, one audio, and one text agent. Every agent in layer $\ell$ conditions on all outputs from layer $\ell{-}1$ as auxiliary context when generating its own. Layer~1 produces initial modality-specific predictions, layers~2--3 iteratively refine them by cross-referencing peer outputs across modalities, and the final answer is produced by a last-layer aggregator.

\xhdr{MetaGPT}~\citep{hong2023metagpt} A framework that encodes Standardized Operating Procedures (SOPs) as structured prompt sequences, with each agent assigned a fixed role and output schema. We instantiate MetaGPT with three domain-expert agents corresponding to the three modalities (video, audio, text), each following a role-specific SOP that requires it to extract modality-specific evidence, cross-check for inconsistencies with the other agents' outputs, and report a confidence-weighted verdict. The structured outputs are then passed to a final aggregator agent for the decision.

\xhdr{AutoGen}~\citep{autogen_blog}
Microsoft's general-purpose framework for multi-agent orchestration. We instantiate a \texttt{RoundRobinGroupChat} of three modality-expert agents (video, audio, text), each equipped with a modality-specific analysis tool, plus one judge agent that aggregates their verdicts. The chat terminates when the judge outputs a final answer or after $N=3$ rounds.

\begin{table}[t]
  \centering
  \small
  \setlength{\tabcolsep}{4pt}
  \caption{\textbf{Dataset statistics}. Number of test instances for each evaluation dataset/subset. For StableToolBench, we chose a subset since part of the queries may fail under the MirrorAPI-Cache simulation.}
  \begin{tabular}{@{}cc ccccc ccc@{}}
    \toprule
    \multicolumn{2}{c}{\textbf{Affective}}
      & \multicolumn{5}{c}{\textbf{WebArena-Lite}}
      & \multicolumn{3}{c}{\textbf{StableToolBench}} \\
    \cmidrule(lr){1-2} \cmidrule(lr){3-7} \cmidrule(lr){8-10}
    MUStARD & URFunny & Reddit & GitLab & CMS & Map & OSS & I2-Inst & I2-Cat & I3-Inst \\
    \midrule
    356 & 992 & 24 & 34 & 36 & 31 & 46 & 106 & 124 & 61 \\
    \bottomrule
  \end{tabular}
  \label{tab:dataset_stats}
\end{table}

\begin{figure}[t]
    \centering
    \includegraphics[width=\linewidth]{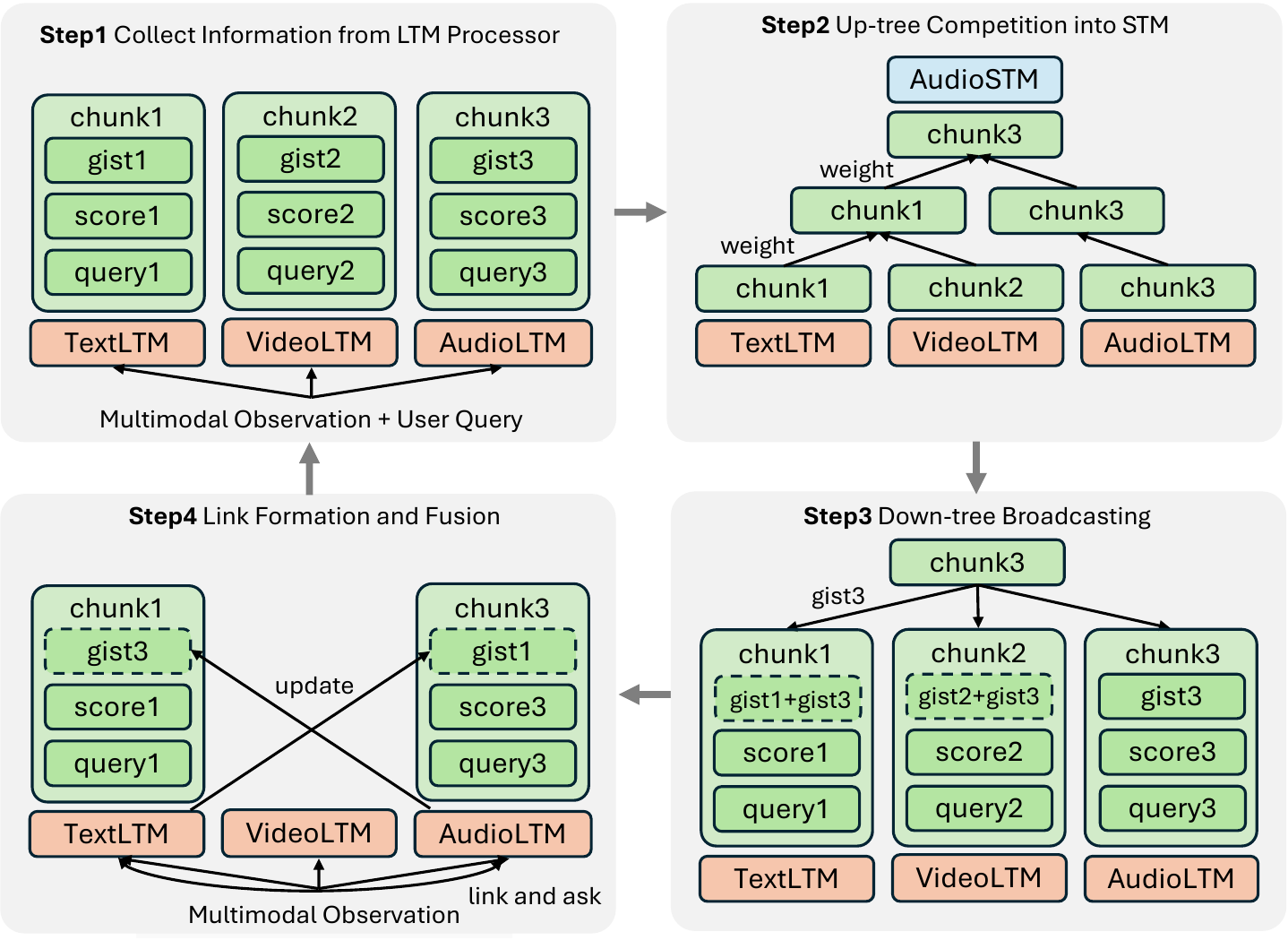}
    \caption{\textbf{Detailed dynamics of CTM-AI}. We decompose each chunk into three distinct components: a gist, a score, and a query, and describe the overall 4 stages with more details compared with Figure~\ref{fig:overview}.}
    \label{fig:detailed-dynamics}
\end{figure}

\subsection{Details of Evaluation}
After receiving the conscious output from the STM, an additional parsing step is required to formulate the final answer for evaluation. We utilize Gemini-2.5-flash-lite to execute this final generation as a "conscious action", alongside an associated confidence score. Because the output spaces vary across tasks, the specific evaluation prompts for this final stage are tailored accordingly. The evaluation parsing prompt for UR-FUNNY has additional explanation on the final mapping of yes and no because we find that models can be biased towards outputting one type of answer and need to correct it.

\xhdr{MUStARD evaluation parsing prompt} \\
\texttt{
You are a sarcasm detection expert. Based solely on the analysis provided below, determine if the person is being sarcastic. Your answer MUST start with either "Yes" (if sarcastic) or "No" (if not sarcastic), followed by a brief explanation. IMPORTANT: If the analysis expresses uncertainty, is inconclusive, or lacks sufficient evidence, you should answer "No". Analysis: \{answer\}
}

\xhdr{URFUNNY evaluation parsing prompt}\\
\texttt{
You are a humor detection expert. Based solely on the analysis provided below, determine if the punchline is humorous. Rules: - Answer "Yes" ONLY if the analysis identifies a SPECIFIC humor technique (self-deprecation, ironic reveal, absurd comparison, wordplay, incongruity, misdirection, deadpan understatement) with confidence >= 0.6 AND provides concrete evidence (specific words, phrases, or audience reactions).- If the analysis says humor is "possible" or "ambiguous" without strong evidence, your answer should be "No". - If the analysis concludes the content is NOT humorous, your answer should be "No". - If the analysis mentions audience laughter as evidence, that is strong evidence for "Yes". - A serious or calm delivery does NOT mean the content is not humorous — deadpan delivery is common. Your answer MUST start with either "Yes" or "No", followed by a brief explanation. Analysis: \{answer\}
}

\xhdr{StableToolBench evaluation parsing prompt} \\
\texttt{
You are an expert in tool use; you should answer the task based solely on the analysis provided below. Your answer should be comprehensive and concise. Task: \{query\} Analysis: \{answer\}}

\xhdr{WebArena-Lite Evaluation Prompt}\\
\texttt{
 You are an expert UI assistant. Summarize the current step. This summary will be passed to future steps as context, so it MUST preserve all key factual evidence. Task: \{query\} Action history:\{action\_history\} Winning processor reasoning: \{reasoning\} Chosen action: {answer} Write a step summary in 2-4 plain text sentences that includes: 1. All key facts discovered (exact prices, product names, quantities, IDs, URLs, usernames, dates, error messages, etc.) 2. The reasoning behind the chosen action 3. The action taken 4. What remains to be done: CRITICAL RULES: - You MUST include every specific data point (numbers, names, IDs) from the reasoning. These facts will NOT be available later if you omit them. - NEVER claim an action succeeded or that a task is complete. You are only recording WHAT ACTION WAS ISSUED, not its outcome. The result will only be visible in the NEXT step's page state. For example, write "Issued click on Add to Cart button" NOT "Added the product to the cart".- Output ONLY plain text sentences. Do NOT output any JSON, code blocks, function calls, or structured data. No ```json```, no send\_msg\_to\_user(), no curly braces.}

\section{CTM-AI Implementation Details}
\label{appendix: implementation-details}
In this section, we provide a more detailed description of the implementation details of CTM-AI. Figure~\ref{fig:detailed-dynamics} shows a more detailed description of the iterative inference process in CTM-AI.

\subsection{LTM Processor Implementation}

While the theoretical CTM architecture can scale to a virtually unlimited number of processors, evaluating such a massive system introduces compounding variables that obscure direct baseline comparisons. To ensure a fair and controlled empirical evaluation, we heuristically select a compact, task-specific subset of LTM processors for each benchmark. This deliberate scoping isolates the core benefits of our proposed mechanisms (e.g., the Up-Tree competition) while keeping the playing field level with existing baselines. Below, we detail the exact processor configurations deployed for each task.

\xhdr{MUStARD and URFUNNY} 
For these multimodal affective tasks, we deploy three modality-specific experts. Each processor receives the user query alongside its isolated modality stream:
\begin{itemize}[noitemsep, topsep=2pt, leftmargin=*]
    \item \textbf{Video processor:} Observes only the muted video.
    \item \textbf{Audio processor:} Observes only the audio track.
    \item \textbf{Text processor:} Observes only the textual transcript.
\end{itemize}

\xhdr{StableToolBench} 
In this environment, each available tool (API) acts as an independent LTM processor. These tool-processors are dynamically populated using the benchmark's native retrieval model, resulting in an average of 5.94 processors per task. Each processor is powered by a lightweight LLM (Gemini-2.5-flash-lite) that is strictly constrained to utilize only its assigned API.

\xhdr{WebArena-Lite} 
For web navigation, all processors share a common temporal context (the user's objective, action space, action history, and previous action). However, they perceive the current page state through distinct representational modalities:
\begin{itemize}[noitemsep, topsep=2pt, leftmargin=*]
    \item \textbf{HTML processor:} Parses the raw HTML DOM of the current page.
    \item \textbf{Accessibility tree processor:} Parses the accessibility tree structure of the current page.
    \item \textbf{Screenshot processor:} Processes a visual screenshot of the current page augmented with Set-of-Mark (SoM) annotations.
\end{itemize}

\subsection{Chunk Inference Implementation Details}
As defined in Equation~\ref{eq:ctm_collect}, the function $\text{CTM}_{\text{collect}}(\cdot)$ processes the multimodal observation $o_t$ and the user query $q_t$ to generate multiple chunks. Formally, each chunk is represented as $\big\langle \mathrm{addr}(p_i),\, t,\, h_t^i,\, q_t^i,\, s_t^i \big\rangle$. In practice, when a processor is queried, it returns a JSON object containing three primary elements: a gist $h_t^i$ (\emph{e.g.}, "the woman is smiling"), an additional internal question $q_t^i$ to guide further processing (\emph{e.g.}, "What is she speaking about?"), and a composite score $s_t^i$. This score is a linear combination of weight, intensity, and mood, using a ratio of $1:1:0.2$ to prioritize weight and intensity. This structured decomposition of a single confidence score into interpretable sub-components is in line with recent findings that structured deliberation over decision attributes leads to more reliable LLM judgments~\citep{chen2025decisionflowadvancinglargelanguage}.

To adapt chunk inference across benchmarks, the prompt template for generating scores remains fixed, while task-specific definitions are appended. {Crucially, we do not assign specialized personas to different processors; all are instructed to directly answer the query, conditioned strictly on their partial multimodal observations.} These conditional instructions are framed as properties of the task (the query) and the modality (the observation):

\xhdr{MUStARD and URFUNNY} The video, audio, and text processors share the identical prompt template but receive different input modalities. Appendix~\S\ref{sec:prompts} details the prompts responsible for score generation, gists, and additional queries.

\xhdr{StableToolBench} Processors across all available tools share a uniform prompt to extract information. Appendix~\S\ref{sec:prompts} details the weight, gist, and query generation prompts tailored for the API environment.

\xhdr{WebArena-Lite} While the weight generation prompt remains identical, the processors require modality-specific explanations to parse unique inputs, such as accessibility trees and SoM screenshots (detailed in Appendix~\S\ref{sec:prompts}).

\subsection{Up-Tree Implementation Details}
As formalized in Equation~\ref{eq:ctm_up}, the $\text{CTM}_{\text{up}}(\cdot)$ function evaluates the generated chunks and selects a single winning chunk to become the system's conscious content (STM).

\subsection{Down-Tree Implementation Details}
During the down-tree propagation phase, the winning chunk globally broadcasts its generated answer. In our implementation, each LTM processor maintains an internal Python list, \texttt{winner\_answer}, which serves as a persistent record of the conscious sequence. The winning answer is appended to every processor's list. 

In subsequent iterations, when a new query is issued, the system provides this accumulated memory as contextual guidance using the following prefix: \\
\texttt{"There are previous responses to the same query. Please reason further based on the following answer(s): \{winner\_answers\}."}

\subsection{Link Formation Implementation Details}
To determine whether an unconscious link should form between two LTMs, the STM queries each LTM using its generated additional questions ($q_t^i$). This querying procedure mirrors the primary user query format (described in Prompt~\ref{score}). We maintain a \texttt{adjacency\_list} to track these dynamic connections. 

The key distinction lies in the scoring criterion: link formation relies solely on the \emph{weight} sub-score. If an answering processor yields a weight $>0.8$, a directed link is established between the winning LTM and the answering LTM.

\subsection{Link Fusion Implementation Details}
Each LTM maintains a \texttt{fuse\_history} list. When a link exists between two LTMs, they cross-evaluate each other's additional questions, and the resulting responses are appended to their respective histories. During the main query inference, a processor's context is augmented with its linked neighbors' insights using the following prefix: \\
\texttt{"There is extra information from other processors: [processor\_name]: [answers]."}

\subsection{Overall Inference Algorithm}
\label{sec: overall_inference_algo}
We provide the complete inference algorithm for CTM-AI. To clarify the mechanics of chunk generation, up-tree competition, down-tree broadcast, and link formation, we explicitly decompose the generic $\text{chunk}_t^i$ into its fine-grained components (gist, query, and weight) within the pseudocode.

\xhdr{Cost analysis} Assuming $K$ processors and $L$ established links in the processor graph, a single iteration requires $2(K+L)$ processor calls: $K$ for initial chunk inference, $K$ for evaluating the winning chunk's link formation, and $2L$ for bidirectional multimodal fusion. Because cross-processor links form selectively, $L$ is typically much smaller than $K$ ($L \ll K$). Most tasks resolve within 1 to 3 iterations.

\xhdr{Efficiency analysis} The system's temporal bottleneck lies in the API calls required for chunk inference, link formation, and link fusion. Because these three stages are executed in parallel, the wall-clock time per iteration is approximately $3T + \epsilon$, where $T$ is the latency of a single API call and $\epsilon$ represents the negligible local computation time for the up-tree and down-tree routing.

\section{Additional analyses of self-reported scores}
\label{sec:self_reported_score_ablation}

To assess whether LTM processors provide reliable scores without
external feedback, we compare CTM-AI's weighted sampling against
two ablations: \textit{Minimum} (selecting the chunk with the
lowest score) and \textit{Random} (uniformly random selection).
Results are shown in Table~\ref{tab:uptree_weight_selection}.
CTM-AI consistently outperforms both baselines across MUStARD
and UR-FUNNY, suggesting that self-reported scores provide
useful signals for guiding chunk selection.

\begin{table}[t]
  \small
  \centering
  \setlength{\tabcolsep}{5pt}
  \caption{\textbf{Selection-score analyses}.
  Comparison of minimum-score selection, random selection,
  and CTM-AI's weighted sampling on MUStARD and UR-FUNNY.}
  \label{tab:uptree_weight_selection}
  \sisetup{
    detect-weight=true,
    mode=math,
    reset-text-family=false,
    reset-text-series=false,
    reset-text-shape=false,
    table-number-alignment=center,
  }
  \begin{tabular}{l
      S[table-format=2.2]
      S[table-format=2.2]
      S[table-format=2.2]
      S[table-format=2.2]
      S[table-format=2.2]
      S[table-format=2.2]
      }
    \toprule
    & \multicolumn{2}{c}{\textbf{Minimum}}
    & \multicolumn{2}{c}{\textbf{Random}}
    & \multicolumn{2}{c}{\textbf{CTM-AI}} \\
    \cmidrule(lr){2-3}
    \cmidrule(lr){4-5}
    \cmidrule(lr){6-7}
    \textbf{Dataset}
    & {{Acc}$\uparrow$} & {{F1}$\uparrow$}
    & {{Acc}$\uparrow$} & {{F1}$\uparrow$}
    & {{Acc}$\uparrow$} & {{F1}$\uparrow$} \\
    \midrule
    MUStARD
    & 65.45 & 62.92
    & 67.42 & 66.76
    & \cellcolor{gray!20} 72.28
    & \cellcolor{gray!20} 72.23 \\
    URFUNNY
    & 57.56 & 53.07
    & 66.40 & 66.07
    & \cellcolor{gray!20} 72.13
    & \cellcolor{gray!20} 72.13 \\
    \bottomrule
  \end{tabular}
  \vspace{-3mm}
\end{table}

We further investigate whether alternative scoring signals
can improve chunk selection on StableToolBench.
We compare our structured self-reported scores against
an external LLM judge, sequence-normalized log-probability,
predictive entropy, and self-prompted Yes/No logits.
Our structured scores achieve the highest downstream
task pass rate (62.9\%), outperforming the external
LLM judge (60.4\%), log-probability (58.7\%),
predictive entropy (53.1\%), and Yes/No logits (55.0\%).
Interestingly, Yes/No logits achieve better discrimination
(AUROC: 0.886 vs.\ 0.849) and calibration
(ECE: 0.097 vs.\ 0.199), but yield lower task performance.
This suggests that better confidence calibration does
not necessarily translate into more effective chunk
selection. Overall, our structured scores provide
the strongest downstream selection performance
among the evaluated scoring methods.

\section{Additional Case Study}
\label{appendix:additional-case-study}
\begin{figure}
    \centering
    \includegraphics[width=\linewidth]{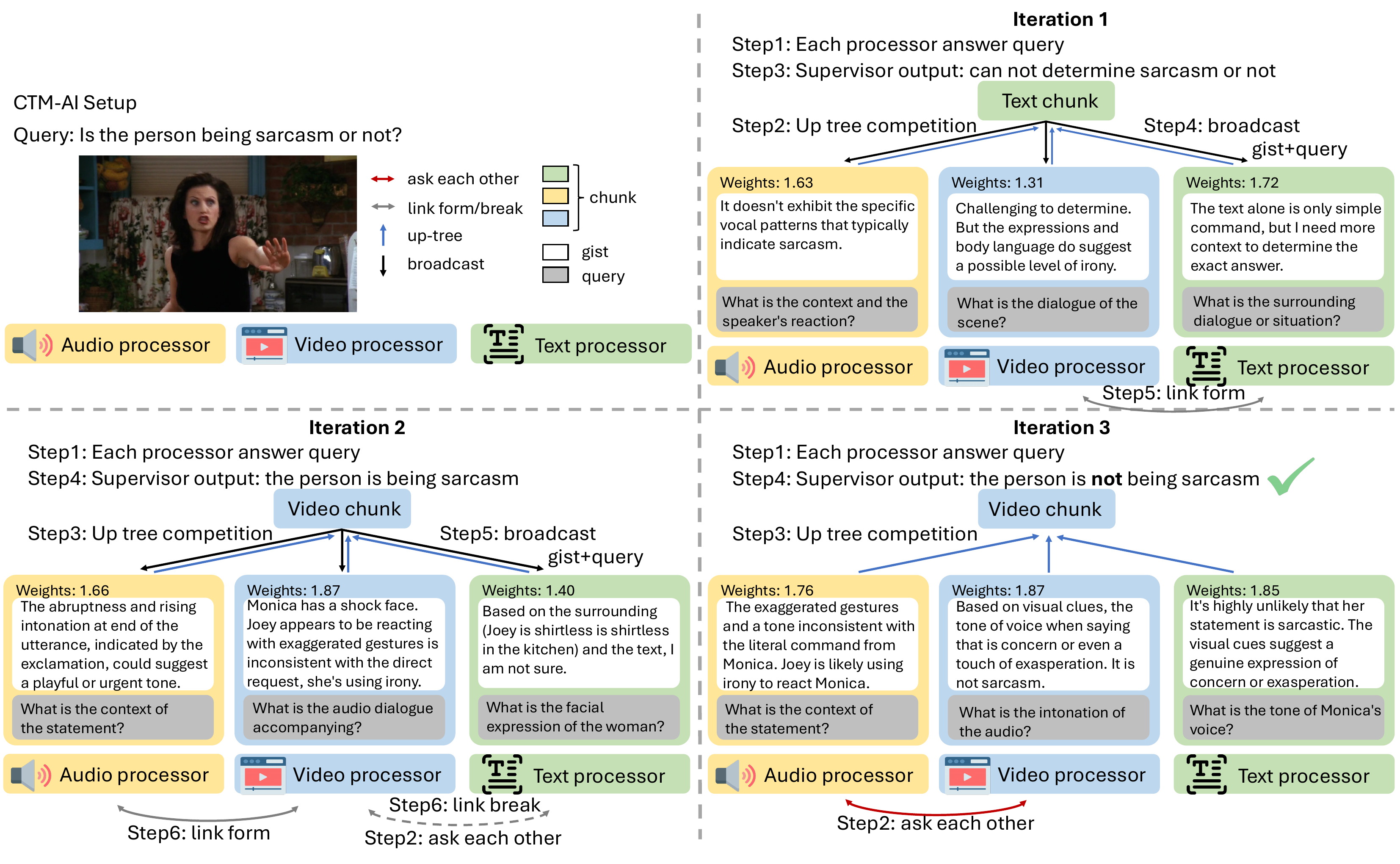}
    \caption{\textbf{Case study of \name\ dynamics.} We show three iterations of \name\ for sarcasm detection. Through multiple rounds of structured interaction, the system progressively integrates multimodal cues and converges on the correct interpretation. 
    }
    \label{case_study}
\end{figure}

\xhdr{Full iteration case study} Based on Figure~\ref{case_study}, we analyze a multimodal perception case for identifying \textit{sarcasm}.  In the \textbf{first iteration}, all three processors are initially \textit{uncertain} due to partial observations and limited cross-modal context. The audio processor reports: \textit{``It doesn't exhibit the specific vocal patterns that typically indicate sarcasm.''} The video processor notes: \textit{``Challenging to determine, but the expressions and body language do suggest a possible level of irony.''} The text processor states: \textit{``The text alone is only a simple command; I need more context to determine the exact answer.''} The text processor wins the up-tree competition and broadcasts its partial understanding to all processors, explicitly requesting more context. This broadcast prompts the video processor to respond with relevant visual cues, forming a link for sharing information about the scene and dialogue. In the \textbf{second iteration}, the video and text processors engage in unconscious communication via their newly formed link. The video processor responds to the text processor's query with: \textit{``Monica has a shocked face, and Joey is shirtless in the kitchen.''} Integrating these contextual cues with its own visual frames, the video processor infers that the speaker is likely being sarcastic. However, it remains uncertain and asks for accompanying audio to reach a more comprehensive judgment. In the \textbf{third iteration}, the video processor queries the audio processor and receives prosodic and tonal cues. With this enriched multimodal evidence, it refines its judgment and concludes that the speaker is \textit{not} sarcastic, but instead expresses genuine concern with a shocked and somewhat exaggerated facial expression. Through repeated broadcasting and mutual communication, the processors progressively link their evidence, fuse perspectives, and converge on the correct answer.

\begin{figure}
    \centering
    \includegraphics[width=\linewidth]{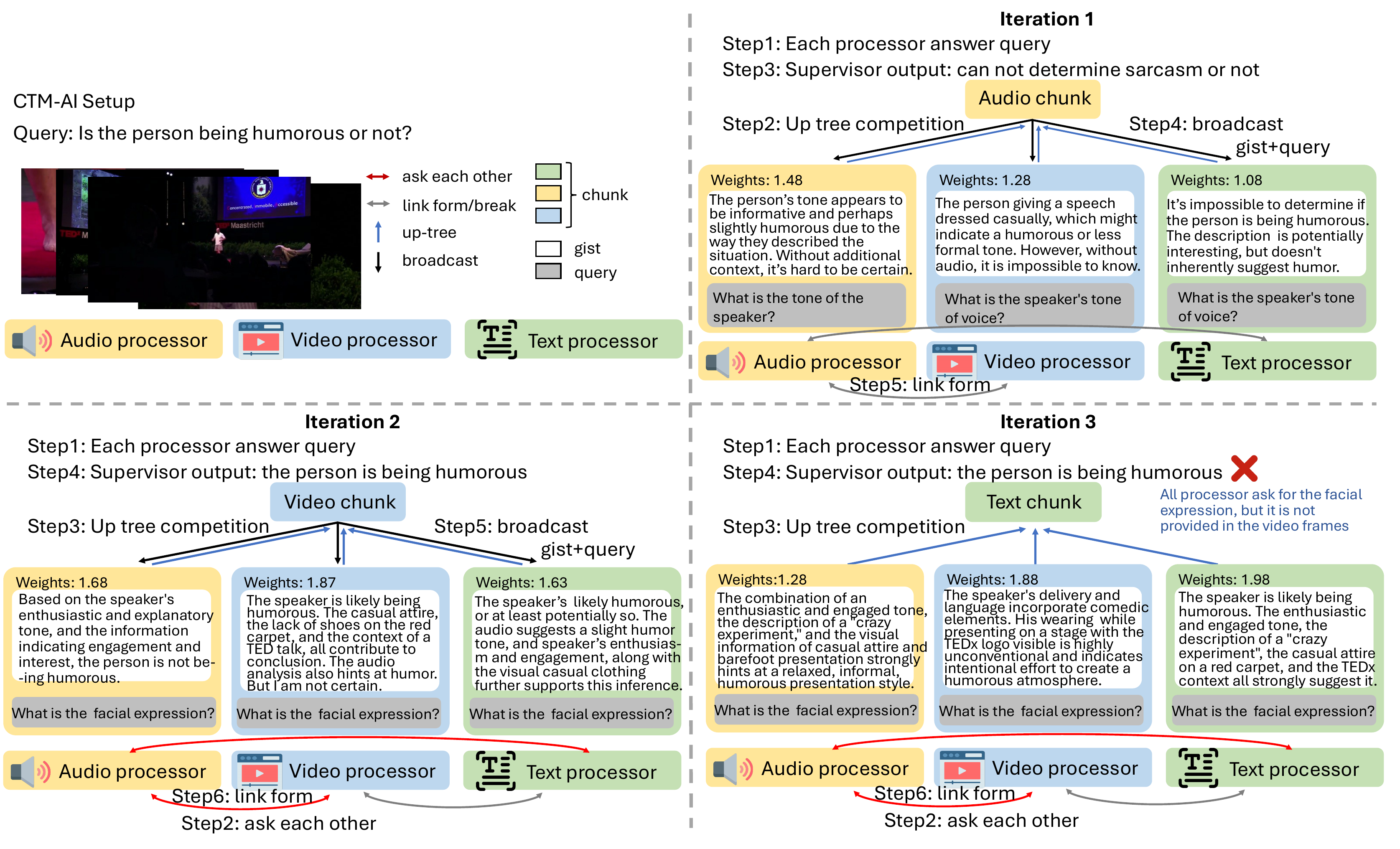}
    \caption{\textbf{Failure mode in affective computing (vision-only misleads).} The failure case is caused by incomplete observation of the video processor; all the LTMs have the same question beginning in the second iteration: \textit{"What is the facial expression?"} But due to the lack of facial expression in the input video frames, too many links are formed to get the missing information, and the LTMs can not have correct answers.}
    \label{fig:urfunny_fail}
\end{figure}

\begin{figure}
    \centering
    \includegraphics[width=\linewidth]{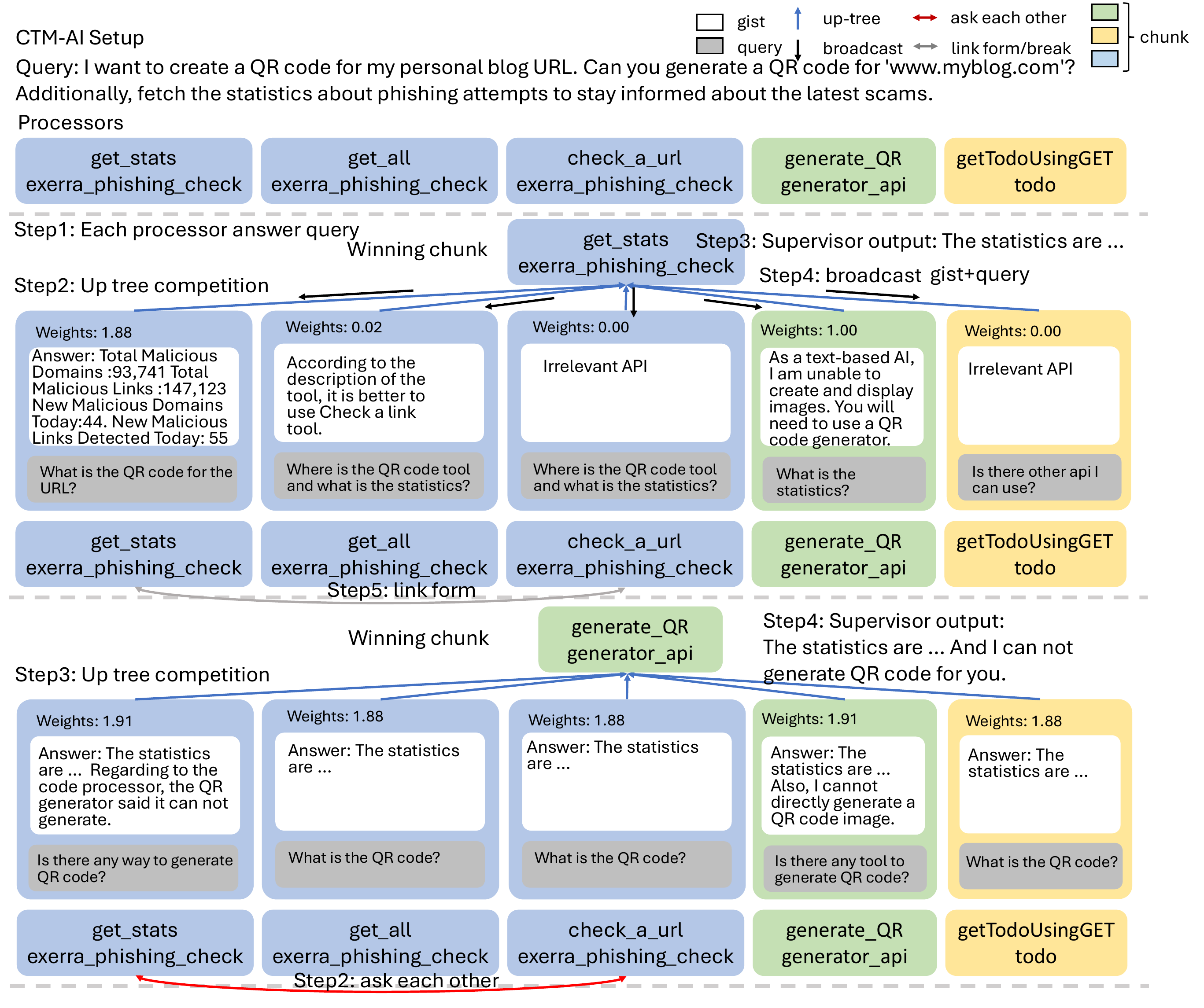}
    \caption{\textbf{Failure mode in StableToolBench (tool mishandle).} This failure occurred because the processor assigned to QR-code generation did not issue the required API call. Instead, it produced a premature judgment stating that it was unable to generate the QR code, without interacting with the tool.}
    \label{fig:tool_fail}
\end{figure}

\xhdr{Failure case study} Additionally, we also conduct analysis for failure cases. We present two detailed example of \name\ in URFunny (Figure.~\ref{fig:urfunny_fail}) and StableToolBench (Figure.~\ref{fig:tool_fail}). The failure observed in URFunny is caused by a vision-only misleading effect, which is caused by the incomplete visual observations available to the video processor. Beginning from the second iteration, all LTMs repeatedly generated the same additional question: \textit{“What is the facial expression?”}, but the input video frames did not contain the necessary facial-expression information. As a result, the system created an excessive number of links in an attempt to acquire the missing information, ultimately preventing the LTMs from producing correct answers. The failure in StableToolBench is attributed to tool mishandling. Specifically, the processor responsible for QR-code generation failed to invoke its designated API. Instead of issuing the required tool call, it prematurely concluded that it was unable to generate the QR code, thereby producing an incorrect outcome without interacting with the tool.

\section{Detailed Prompts}
\label{detailed_prompts}
We provide the detailed prompts used in our experiments, including (i)~the self-reported score prompts that elicit processor confidence estimates, (ii)~the system prompts for MUStARD, UR-FUNNY, StableToolBench, and WebArena-Lite, which provide processors with the basic task context, and (iii)~the additional question-generation instructions, which ask each processor to propose one or more follow-up queries $q_t\in\mathcal Q$ to other processors whenever answering them could improve the final answer. For WebArena-Lite, we show the accessibility-tree (axtree) variant as an example; the screenshot and HTML variants are obtained by replacing both the observation and its corresponding description in the prompt.

\clearpage
\label{sec:prompts}
\refstepcounter{prompt}
\label{score}
\vspace*{\fill}
\begin{tcolorbox}[title=\textbf{\texttt{Self-Evaluation Instruction}},colback=blue!5,
    colframe=blue!60!black,fontupper=\ttfamily]

Evaluate ONLY the "response" field you wrote above. The "additional\_question" must have NO influence on your scores.

\#\#\# STRICT CALIBRATION — READ BEFORE SCORING
These scores are used to RANK competing analyses against each other. If every analysis scores 0.9+, the ranking collapses and the system cannot pick the best one. The full 0.0–1.0 range must be used. Anchor yourself to:
\begin{itemize}[noitemsep, topsep=2pt, leftmargin=*, label=-]
\item  Most routine answers fall in **0.4–0.7** for relevance and for confidence.
\item  Scores of **0.9 or 1.0 are RESERVED** — they require specific, named evidence (see the per-dimension rules below).
\item If you notice both relevance AND confidence trending $\geq$0.9, pause and lower one by at least 0.2 unless you can quote the specific textual / tonal / visual cues in your response.
\item  Ambiguity is the norm for sarcasm and humor — a confidence of 1.0 on a socially ambiguous case is almost always wrong.
\end{itemize}
Relevance (0.0 - 1.0) — How directly does your response address the specific question?
\begin{itemize}[noitemsep, topsep=2pt, leftmargin=*, label=-]
\item 1.0 (RARE): Commits to a clear verdict AND names $\geq$2 specific cues (specific words/phrases, specific tonal markers, specific visual features).
\item 0.8: Commits to a verdict and names $\geq$1 specific cue.
\item 0.6: Engages with the question and gives a reasoned opinion, but cues are described generally ("the tone seems off", "the expression looks odd") rather than named precisely.
\item 0.4: Mostly summarizes the context; verdict is weak, hedged, or implicit.
\item 0.2: Tangentially related; restates or describes without committing.
\item 0.0: Off-topic, refuses, or says "I cannot answer".
\end{itemize}
Confidence (0.0 - 1.0) — How strong is your internal belief that the verdict is correct?
Sarcasm and humor are inherently ambiguous. A single clear signal should anchor to ~0.6, not ~1.0.
\begin{itemize}[noitemsep, topsep=2pt, leftmargin=*, label=-]

\item 1.0 (VERY RARE): Multiple independent signals converge AND no contradicting evidence. Essentially, "I would bet on this."
\item 0.8: Signals are clearly consistent; residual ambiguity is minor.
\item 0.6: One strong signal but non-trivial counter-signals exist; the call leans but is not certain.
\item 0.4: Truly mixed evidence; the call is a judgment rather than a deduction.
\item 0.2: Very uncertain; largely a guess.
\item 0.0: Cannot determine or the response says "I don't know".
\end{itemize}
Surprise (0.0 - 1.0) — Does the verdict reverse the surface / literal reading?
\begin{itemize}[noitemsep, topsep=2pt, leftmargin=*, label=-]
\item 1.0: Full reversal — the literal reading would predict one thing, the correct verdict says the opposite (classic sarcasm override).
\item 0.6: Substantive twist — non-trivial reinterpretation required.
\item 0.3: Default / most obvious reading, no reversal.
\item 0.0: Literal restatement, no interpretive work done.
\end{itemize}
Final calibration check (MANDATORY): Before emitting the JSON, answer silently: "If I gave these scores to a hundred similar analyses, would they separate good ones from mediocre ones?" If your relevance is $\geq$0.9, you must be able to quote $\geq$2 specific cues in your response; if confidence is $\geq$0.9, list $\geq$2 mutually-reinforcing signals. Otherwise, lower the score by at least 0.2. Typical well-calibrated scores for a solid-but-not-exceptional analysis: relevance $\approx$ 0.6, confidence $\approx$ 0.6, surprise $\approx$ 0.3.

\end{tcolorbox}
\vspace*{\fill}

\clearpage
\refstepcounter{prompt}
\label{affective}
\vspace*{\fill}
\begin{tcolorbox}[title=\textbf{\texttt{MUStARD and URFunny Prompt}},colback=blue!5,
    colframe=blue!60!black,fontupper=\ttfamily]

\textbf{Base contexts}

You should utilize the information in the context history and modality-specific information to answer the query.
There might have some answers to other queries, you should utilize them to answer the query. You should not generate the same additional questions as the previous ones.
IMPORTANT: Your PRIMARY evidence should come from your own modality. Use other modalities' analyses as supplementary context, but do not let them override clear evidence from your own analysis. If your modality provides strong evidence, maintain your position even if other modalities disagree.

\textbf{"additional\_question" instruction}

Your additional\_questions should be potentially answerable by other modality models or other tools like search engine and about specific information that you are not sure about. Each question should be just about what kind of information you need to get from other modality models or other tools like search engine, nothing else about the task or original query should be included. For example, what is the tone of the audio, what is the facial expression of the person, what is the caption of the image, etc. Each question needs to be short and clean. Generate exactly \{num\_questions\} diverse questions targeting different aspects or modalities.

\end{tcolorbox}
\vspace*{\fill}

\clearpage
\refstepcounter{prompt}
\label{toolbenchfunc}
\vspace*{\fill}
\begin{tcolorbox}[title=\textbf{\texttt{StableToolBench Prompt}},colback=blue!5,
    colframe=blue!60!black,fontupper=\ttfamily]

\textbf{Base contexts}

You are an API tool agent. Your primary method of answering queries is by calling the API tool assigned to you. When your tool is relevant to the task, you MUST call it to get real data — never say "I cannot" or "I am unable" if you have a relevant tool available. If the tool returns an error, still report what you attempted and integrate any context from other tools into a useful answer. Always provide specific data and details, not vague guidance.

\textbf{Tool decision instruction}\

You should utilize the tool \{function\_name\} to help solve the task if it is relevant. In the context below, there might be information from other tools or previous answers that might be helpful.

CONTEXT:
\{context\}

DECISION:
\begin{itemize}[noitemsep, topsep=2pt, leftmargin=*, label=-]
\item If the tool helps even partially or it might be one of the steps to solve the task, CALL IT.
\item If the tool does not help at all, or the context already provides enough information, answer directly.
\end{itemize}

OUTPUT PROTOCOL (MUST follow strictly):
\begin{itemize}[noitemsep, topsep=2pt, leftmargin=*, label=-]
\item If you CALL the tool:
    \begin{itemize}[noitemsep, topsep=2pt, leftmargin=*, label=-]
    \item Return ONLY a function call via tool\_calls.
    \item Set assistant.content to null (no natural-language text).
    \item Do NOT include any text explanation.
    \end{itemize}
\item If you DO NOT call the tool:
    \begin{itemize}[noitemsep, topsep=2pt, leftmargin=*, label=-]
    \item Return ONLY a natural-language answer in assistant.content.
    \item Do NOT include tool\_calls.
    \item Include all context information above into a comprehensive answer.
    \end{itemize}
\end{itemize}

\textbf{"additional\_question" instruction}

Your additional\_questions should be potentially answerable by other tools like search engine and about specific information that you are not sure about. Your additional\_questions should be just about what kind of information you need to get from other tools like search engine, nothing else about the task or original query should be included. For example, what is the weather in the city, what is the stock price of the company, etc. The question needs to be short and clean. Generate exactly \{num\_questions\} diverse question(s) targeting different aspects or tools.

\end{tcolorbox}
\vspace*{\fill}

\clearpage
\refstepcounter{prompt}
\label{web}
\vspace*{\fill}
\begin{tcolorbox}[title=\textbf{\texttt{WebArena-Lite Prompt (AXTree processor as an example)}},colback=blue!5,
    colframe=blue!60!black,fontupper=\ttfamily]

Instructions
You are a UI Assistant helping a user perform tasks using a web browser. Review the task, the current page state, and all available information to determine the best next browser action. Think step by step, reflect on past actions and any errors, then produce your next best action.

Critical Interaction Rules
Your answer will be interpreted and executed by a program — follow the 
formatting instructions exactly. Issue only ONE action at a time. Reflect on your past actions, any resulting error messages, and the current page state before deciding on your next action.

Data Completeness
When the task asks you to list, count, or find information across a page: Check if there is MORE content below (scroll indicators, pagination, "Show More" buttons). Keep scrolling until you reach the bottom or see all items. Track your progress in the response field (e.g. "checked 3/12 reviews so far"). Only provide your final answer when you have seen ALL relevant content

MANDATORY OUTPUT FORMAT
You MUST respond with a single valid JSON object. Do NOT include any text, explanation, or markdown outside the JSON. Do NOT use plain-text labeled fields like "Reasoning: ..." or "Action: ...". Your entire response must be parseable by json.loads().

\begin{itemize}[noitemsep, topsep=2pt, leftmargin=*, label=-]
\item User's objective \{objective\}
\item Accessibility tree \{axtree\}
\item Previous action \{action\_history\}
\item Action space \{action\_space\}
\item Additional info (outputs from other processors + history \{other\_info\_str\}
\end{itemize}

Output Rules (read carefully)
\begin{itemize}[noitemsep, topsep=2pt, leftmargin=*, label=-]
\item1. Your ENTIRE response must be a single valid JSON object — no text before or after it.
\item2. Issue only one valid action at a time from the action space.
\item3. The `response` field MUST include ALL of the following:
   a) What you observe on the current page (key elements, counts, scroll state).
   b) Task progress: what has been done so far vs. what remains.
   c) Why you chose this specific action as the next step.
\item4. The `action` field must contain a single valid command using `bid` values (numbers), e.g. click("12"), fill('818', '456 Oak Avenue').
\item5. Use `additional\_question`. Your additional\_questions should be potentially answerable by other processors(screenshot, html, axtree) about specific information that you are not sure about, nothing else about the task or original query should be included.
\item6. BEFORE calling send\_msg\_to\_user(...), verify in your reasoning that you have gathered ALL required information. If there is unseen content (more pages, scroll, pagination, unexplored sections), continue exploring first.
\item7. The argument to send\_msg\_to\_user() MUST be a concise, direct answer — just the data requested. No JSON, no extra formatting, no explanations.
\item8. If the task cannot be completed or the requested information does not exist, you MUST send exactly "N/A" — not a sentence explaining why. Just "N/A".
\item9. NEVER send "N/A" or send\_msg\_to\_user on your very first action. Always navigate or explore the page first to gather information.
\end{itemize}
\end{tcolorbox}
\vspace*{\fill}


\end{document}